\documentstyle[epsfig]{mn}

\newcommand{\be}{\begin{equation}}
\newcommand{\ee}{\end{equation}}
\newcommand{\gs}{\raisebox{-.8ex}{$\buildrel{\textstyle>}\over\sim$}}
\newcommand{\ls}{\raisebox{-.8ex}{$\buildrel{\textstyle<}\over\sim$}}

\newcommand{\anrev}{{\it ARA\&A, }}
\newcommand{\apj}{{\it ApJ, }}

\newcommand{\icar}{{\it Icarus, }}
\newcommand{\mnr}{{\it MNRAS, }}
\newcommand{\nat}{{\it Nat, }}

\newcommand{\ana}{{\it A\&A, }}

\newcommand{\MJup}{M_{J}}
\newcounter{pp3}
\addtocounter{pp3}{3}

\title[Migration of protoplanets]
{The Migration and Growth of Protoplanets in Protostellar Discs}
\author[Nelson, Papaloizou, Masset, Kley]{Richard P. Nelson$^1$ \thanks{For preprints email: R.P.Nelson@qmw.ac.uk}, John C.B. Papaloizou$^1$, Fr\'ed\'eric Masset$^1$, \& Willy Kley$^2$ \\
 $^1$ Astronomy Unit, 
 Queen Mary \& Westfield College, Mile End
 Rd, London E1 4NS\\
 $^2$ Theoretisch-Physikalisches Institut,
     Universit\"at Jena, Max-Wien-Platz~1, D-07743 Jena, Germany\\
}

 \date{Received *****; in original form ******}

 \def\LaTeX{L\kern-.36em\raise.3ex\hbox{a}\kern-.15em
         T\kern-.1667em\lower.7ex\hbox{E}\kern-.125emX}

\begin{document}
 \label{firstpage}

 \maketitle

 \begin{abstract}
We investigate the gravitational interaction of a Jovian mass protoplanet
with a gaseous disc with aspect ratio and kinematic
viscosity  expected for the protoplanetary disc from which it formed.
Different  disc surface density distributions have been investigated.
We focus on the  tidal interaction with the disc
with  the consequent gap formation and orbital migration of the protoplanet.
Nonlinear hydrodynamic simulations are employed using three 
independent numerical codes.
 
A principal result is that the direction
of the orbital migration is always inwards and such that the
protoplanet reaches the central star in  a near circular orbit
after a characteristic viscous time scale of $\sim 10^4$
initial orbital periods. 
This was found to be independent
of whether the protoplanet was allowed to accrete mass or not.
Inward migration  is helped  through
the disappearance of the inner disc,
and therefore the positive torque it would exert,
because of accretion onto the central star.
Maximally accreting protoplanets reached about four 
Jovian masses 
on reaching  the neighbourhood of the central star.
Our results indicate that a realistic upper limit for the masses of closely
orbiting giant
planets is $\sim 5$ Jupiter masses, because of the reduced accretion rates
obtained for planets of increasing mass.

Assuming some process such as termination of the inner disc
through a magnetospheric cavity stops the migration,   
the range of masses estimated for a number of close orbiting
giant planets  (Marcy, Cochran, \& Mayor 1999; Marcy \& Butler 1998)
as well as their inward orbital migration can be
accounted for by consideration of disc--protoplanet interactions
during the late
stages of giant planet formation.
\end{abstract}

\begin{keywords} giant planet formation- extrasolar planets-
accretion discs- numerical simulations
 \end{keywords}

\section{Introduction} \label{intro}

The recent discovery of a number of extrasolar giant planets orbiting around
nearby solar--type stars has stimulated renewed interest in the
theory of planet formation. These planetary objects have masses, $m_p$, that
are comparable to that of Jupiter
($0.4 \; {\rm M}_J \; \ls \; m_p \ \ls \; 11 \; {\rm M}_J$), 
have orbital semi-major axes in the range 
$0.04 \; {\rm AU} \; \ls \; a \; \ls \; 2.5 \; {\rm AU}$, and orbital 
eccentricities in the range $ 0.0 \; \ls \; e \; \ls \; 0.67$ 
(Marcy, Cochran, \& Mayor 1999, Marcy \& Butler 1998 and references therein).
It should be noted that
the detection technique of measuring the Doppler shift induced by the
host star's orbital reflex motion only allows the measurement of $m_p \sin{i}$,
where $i$ is the inclination angle of the orbit plane to the line of sight.

It is generally believed that planets form out of the gas and dust 
contained in the discs
that are observed around young T Tauri stars (Beckwith \& Sargent 1996).
This T Tauri disc phase of
a stars life is thought to last on the order of $10^6$--$10^7$ yr, after which
the discs appear to dissipate.
In the standard theory, planet formation occurs 
through a number of key stages. First, the dust grains,
that are initially well mixed with the gas in the disc, undergo coagulative
growth {\em via} binary collisions. Second, as the grains 
continue to grow they
begin to gravitationally settle towards the midplane of the protostellar disc,
forming a dense dust layer in the process. The existence of this dense layer
enhances the rate at which the solid material may combine into larger bodies,
leading eventually to the formation of planetesimals. Third, the 
planetesimals continue to
grow through collisions, possibly aided
by a runaway accretion process (e.g. Lissauer \& Stewart 1993)
 ultimately forming planetary sized objects. These authors estimate
that the time scale for this to occur at 5 AU is $10^5$--$10^6$~yr, although
this  would 
require a high dust to gas ratio in a minimum mass solar nebula.
However, there are many uncertainties in the processes involved
and the effect of  phenomena such as disc--planet interactions
and  orbital migration considered in this paper
have yet to be explored in detail.  The latter might decrease
the time scale by providing increased mobility of protoplanets
in the nebula.

In the  `critical core mass' model
of giant planet formation, the formation
proceeds  through the build--up of a critical
$\sim 15$ M$_{earth}$ solid core, beyond which mass rapid
 gas
accretion occurs leading to
the formation of a gas giant planet.
Because this process must occur prior to the loss of the gas from the disc
({\em i.e.} within $\sim 10^7$ yr)
it is expected that the cores of
gas giant planets should  begin to form beyond a radius of $r \sim 4$ AU, the
so--called `ice condensation radius' or `snow-line'. Beyond this radius
the temperature in the protostellar disc 
falls below the level that allows volatile
materials to condense out into the solid phase and form ices. The presence of
this additional solid material increases the rate at which solid materials
may coagulate into larger objects, and thus shortens the time required to form
the solid cores that are thought to be the precursors to gas giant planets.

If giant planets begin to
 form beyond radii of $\sim 4$ AU from their host stars,
then orbital migration must have occurred in order to explain the existence of
the closely orbiting extrasolar giant planets. The most likely cause of
this orbital migration is through the gravitational interaction between
the embedded protoplanet and the protostellar disc. 
The linearised response of a gaseous disc to the presence of an embedded
satellite has been investigated by a number of authors
(e.g. Goldreich \& Tremaine 1978, 1979, 1980; Lin \& Papaloizou 
1979a,b, 1980, 1993; Papaloizou \& Lin 1984; Ward 1997 and references therein;
Artymowicz 1993a,b).
The perturbation of an accretion disc by a protoplanet
 leads to the excitation of 
spiral density waves at Lindblad resonances which carry with them 
an associated angular momentum flux that is deposited in the disc at 
the location where the waves are damped. The disc orbiting beyond
the position of the planet receives angular momentum from the planet,
whereas the inner disc loses angular momentum to the planet.
In the situation where the tidal torques are greater than the internal viscous
torques in the disc, and the disc response becomes non linear, it is expected 
that an annular gap, or surface density depression, may be formed in the
vicinity of the planet (Papaloizou \& Lin 1984, Lin \& Papaloizou 1993).
This tidal truncation of the protostellar nebula 
 was investigated using non linear numerical simulations by
Bryden {\em et al.} (1999), hereafter BCLNP,
 and Kley (1999a), in order to examine the effect
of gap formation on the mass accretion rate by an embedded
giant protoplanet. The results of these studies indicate that, for physical
parameters typical of protostellar disc models,
gap formation can substantially reduce the accretion rate, leading to
expected planet masses in the range $1 < m_p < 10$ M$_J$, in close agreement
with the observed masses of the extrasolar planets.

The exchange of angular momentum between the planet and the disc leads to
the possibility of the planet undergoing orbital migration if
there exists an imbalance between the torques exerted
by the inner and outer discs (Goldreich \& Tremaine 1980).
The migration that results may be described by one of two different formalisms,
depending on whether the disc response is linear or nonlinear (with a gap 
forming). 

Type I migration occurs when the disc response is linear
and the background surface density profile remains 
essentially unchanged by the interaction.
A natural tendency for the outer disc torques to 
dominate the inner disc torques
results in the inward migration 
of the planet independently of the background 
disc flow, and typically results in rapid migration time scales of 
$10^5$ or $10^4$ yr for a 1 or 10 M$_{earth}$ planet located at 5 AU, 
respectively (Ward 1997).

Type II migration occurs when the disc response becomes non linear and
a clear gap is formed around the vicinity of the planet. In this case,
provided the planet mass is less than or comparable to the local
disc mass with which it interacts, the migration  occurs on the viscous
evolution time scale of the disc. This process was investigated in detail
by Lin \& Papaloizou (1986). When the mass of the planet becomes larger than
the local disc mass, then the inertia of the planet becomes important
in determining the migration rate of the planet. This situation has
been investigated by Syer \& Clarke (1995), and more recently by
Ivanov, Papaloizou \& Polnarev (1999) in the context of  satellite
black holes orbiting in  AGN discs.  

Both type I and type II migration may be important at different
stages of the planet formation process. In particular it is possible that
solid cores undergo type I migration before gas accretion leading
to giant planet formation at radii smaller than 
4 AU (Papaloizou \& Terquem 1999) and possibly in {\it situ} close to
the central star.

Extrasolar giant planets, however, are observed
to orbit over a wide range of radii, some of which must
have been  contained within the original gas disc prior to its dissipation.
Accordingly, in this paper we  shall focus on the situation
where a giant planet  is assumed to have formed at some radius in the disc
and investigate the subsequent type II migration that occurs as a consequence
of disc--protoplanet interaction.

We employ three independent Eulerian hydrodynamic codes
to examine the non linear evolution of a combined star, disc, and  embedded
protoplanet system. The main questions that we wish to address are:
\begin{enumerate}
\item what is the time scale for an initially embedded giant protoplanet
to migrate in to the close vicinity of its host star~?
\item How much mass is the planet able to accrete from the disc during this
time, and how does the mass evolution of the planet affect its orbital
migration rate~?
\item Does the tidal interaction between the disc and planet
cause the growth of orbital eccentricity during the migration~?
\end{enumerate}

The physical parameters that we focus on in this paper are appropriate
to a 1 M$_J$ protoplanet embedded in a minimum mass solar nebula model
containing $\sim 2$ M$_J$ within 5 AU ({\em i.e.} 
the initial orbital radius of the protoplanet). We find that the protoplanet
migrates in towards the host star approximately
on the viscous evolution time of the disc,
independently of the details of the initial conditions of the simulations, or
the numerical code used. For a planet initially located at a radius of
5 AU this translates into a time of $\sim 10^5$ yr for the disc parameters
that we employ. This time is substantially shorter than the expected life times
of protoplanetary discs, and indicates that orbital migration is an important
factor during the formation epoch of all planetary systems.

Assuming maximal accretion,
the estimated final masses of the planets as they approach their host stars
is found to be in the range $ 2.7 \; \ls \; m_p \; \ls \;  5.65$ M$_J$,
depending
on the  details of the calculation. 
These values  fit in well
with the observed mass range of the extrasolar planets.

The orbits of the planets in all calculations were
found to remain essentially circular.
Thus the observed eccentricities of the extrasolar planets are not reproduced
by our current models. These  might be produced if the disc
had a lower viscosity
resulting in wider and deeper gaps than obtained here
or by the perturbing presence of additional planets in the system.
These issues will be the subject of a future investigation.

This paper is organised as follows. In section~\ref{migration}
we present a more quantitative
discussion of gap formation and orbital migration. In section~\ref{discs}
we present a brief discussion of protostellar disc models, and our choice
of disc parameters. This is followed by a discussion of the equations of 
motion, boundary and initial conditions,
physical parameters, and treatment of the protoplanet in 
section~\ref{phys-mod}. We then go on to describe the hydrodynamic
codes that we use in section~\ref{codes}. The results of the calculations are
described in section~\ref{calculations}. There, we discuss
in detail the results of one calculation (our standard run), and
then examine how the results depend on: (1). the presence or not
of an initial gap in the vicinity of the planet; (2). whether the planet is
accreting or non accreting; (3). the numerical resolution; (4). the code
used to perform the calculation. We also present the results of one very long time scale evolution calculation and investigate the
effects of changing the initial density profile.
Finally, in section~\ref{conclusion} we discuss the broader implications
of our results and draw our conclusions.

\section{Orbital migration and gap formation} \label{migration}
The tidal interaction between an accretion disc and an embedded
protoplanet leads to the exchange of angular momentum between them.
The nonaxisymmetric surface  density perturbation of the more
slowly rotating 
disc exterior to the  orbital radius of the planet  produces 
a negative  gravitational torque   acting 
on the planet.  Similarly the more rapidly rotating
 inner disc exerts
a positive torque. Any imbalance between these inner and outer torques
will lead to the orbital migration of the planet (Goldreich \& Tremaine 1980).

 By Newton's third law 
the planet exerts  oppositely directed 
torques on  the inner and outer disc material.
An annular gap may be formed locally in the disc if the magnitudes
of these
torques exceeds
the internal viscous torques (Papaloizou \& Lin 1984).
The presence or absence  of a
gap determines whether the migration is of type II or I, respectively (Ward 1997).
\subsection{Type I Migration} \label{typeI}
Type I migration occurs when the response of the disc to tidal
forcing by an embedded planet is linear. Then no gap is
formed and the  background
surface density profile  is approximately unchanged.

The presence of a planet orbiting in a gaseous disc leads to the
excitation of trailing spiral
density waves at the Lindblad resonances in the disc 
(Goldreich \& Tremaine 1979).
These density waves carry with them an associated angular momentum flux
which is deposited locally in the disc at the location where the waves
damp. The gravitational coupling between the trailing wave pattern
and the planet leads to a torque acting on the planet.

A natural imbalance between the torques acting on the outer
and inner disc
arises because the locations of the outer Lindblad resonances 
tend to be closer
to the planet's
 position than the inner ones, leading to a net inward migration
of the planet.  This sense of migration is insensitive to details of 
the background disc flow (e.g. Ward 1997).

The differential torque induced by the Lindblad resonances may be written
as (Ward 1997)
\be \Delta T \simeq \left(\frac{c_1}{2}\right) q^2 \Sigma r_p^2 (r_p \Omega)^2
\left( \frac{r}{H} \right)^3 \label{ward-torque} \ee
where $q$ is the planet to central star mass ratio $m_p/M_*$, $\Sigma$ is the
surface density of the disc material, $r_p$ is the planet orbital radius,
$\Omega$ is the Keplerian angular velocity, $H/r$ is the disc
aspect ratio, and $c_1$ accounts for the torque imbalance between the two 
sides of
the disc and is expected to scale
as $\sim (H/r)$. The corresponding radial velocity of migration of
the planet is given by
\be \frac{dr_p}{dt} \sim - c_1 q \left(\frac{\Sigma r_p^2}{M_*}\right) 
(r_p \Omega) \left(\frac{r}{H}\right)^3 \label{rdot-I} \ee
This equation predicts that the radial migration will occur at a higher rate 
for a more massive planet, and remains valid until
the disc response becomes non linear
and a gap begins to form.
 A sufficient
 condition for non linearity through shock
formation   is (e.g. Korycansky \& Papaloizou 1996)
\be q > \left( \frac{H}{r} \right)^3 \label{linear-cond} \ee
This corresponds to a planet of mass $\sim 30$ M$_{earth}$ in
a protostellar disc with  a typically expected
aspect ratio $H/r \sim 0.05$ (see Papaloizou \& Terquem 1999 and
references therein).
Accordingly we expect, as is confirmed by
the simulations presented here, that 
type I migration applies to protoplanets
of substantially smaller mass than we consider in this paper.
\subsection{Type II Migration} \label{typeII}
When the disc response becomes non linear a gap forms, inside of which
the planet orbits. The conditions for a gap to 
form
have been discussed by Papaloizou \& Lin (1984), Lin \& Papaloizou (1985, 1993)
and BCLNP.
These are the thermal or shock formation
condition given by Eq.~(\ref{linear-cond}) and the condition
that tidal torques exceed viscous torques which may be written
\be q > \frac{40}{{\cal R}} \label{visc-cond} \ee
where ${\cal R} = r^2\Omega /\nu $ is the Reynolds' number,
and $\nu$ is the kinematic viscosity. 

When there is a gap and  the planet mass is less than or comparable to
the local disc mass with which it gravitationally interacts,
the migration rate of the planet is controlled by the viscous evolution of
the disc, since the planet then behaves as a representative  particle
in the disc.
In this case the migration rate is given by the radial drift velocity of
the gas due to viscous evolution, which for a steady state disc is given by
\be \frac{dr_p}{dt} \sim \frac{3 \nu}{2 r_p}. \label{rdot-II} \ee
 This leads to a migration time 
\be \tau_{mig} \sim \frac{2 r_p^2}{3 \nu}. \label{visc-time} \ee

When the mass of the planet is larger than the characteristic disc mass
with which it tidally interacts, the inertia of the planet becomes
important in slowing down the orbital evolution.
The inertia of the planet acts as a dam against the viscous
evolution of the disc, and can lead 
to a substantial change in the disc structure
in the vicinity of the planet. The coupled disc--planet evolution in this
case has been considered by Syer \& Clarke (1995) and more recently
by Ivanov, Papaloizou \& Polnarev (1999). Using a simple analytical model,
Ivanov {\em et al}. (1999) estimate the migration time of a massive
planet to be
\be \tau_{mig} \sim \left( \frac{1}{10} \frac{M_{d0}}{m_p} \right)^{1/5}
\frac{m_p}{{\dot M}} \label{tmig-Ivanov} \ee
for a disc with  constant $\nu,$
where ${\dot M}$ is the mass accretion rate through the disc
and $M_{d0}$ is the characteristic unperturbed disc mass that would be contained
within the orbital radius.
We can write ${\dot M}=M_{d0}/\tau_{\nu}(r_p)$,
where $\tau_{\nu}(r_p)=2 r_p^2/(3 \nu)$ is the viscous evolution time of the
disc at a distance $r_p$, and $M_{d0}= \pi r_p^2 \Sigma $
leading to 
\be \tau_{mig} \sim \frac{2}{3 \nu} \left( \frac{1}{10 \pi^4 \Sigma^4}
\right)^{1/5} m_p^{4/5} r_p^{2/5}. \label{tau-mig-big-plan} \ee
If we write $\tau_{mig} \sim r_p (dr_p/dt)^{-1},$ then we obtain the
following relation for the migration rate
\be \frac{dr_p}{dt} \sim \frac{3 \nu}{2} (10 \pi^4 \Sigma^4)^{1/5} m_p^{-4/5}
r_p^{3/5}. \label{rdot-Ivanov} \ee
Thus, we see that as the mass of the planet increases, or its orbital radius
decreases, the rate of migration should slow down. The latter effect arises
because the planet interacts with a smaller amount of disc
mass  at smaller radii. This analysis also predicts that
a protoplanet with mass substantially larger than $M_{d0}$
should not increase its mass significantly before
migrating to the centre of the disc. This together
with the reduction of the accretion rate with increasing
protoplanet mass (BCLNP and work presented here)
suggests that the protoplanet mass should also be limited
at about $M_{d0}.$

If we consider the interaction of a Jupiter mass planet initially
at 5~AU  with a minimum mass
solar nebula disc model containing $\sim 2$ Jupiter masses within 5~AU, we see
that the migration rate should initially occur at the viscous evolution rate
of the disc given by Eq.~(\ref{rdot-II}) since $m_p < M_{d0}$.
However, if the planet accretes mass and/or  migrates inwards,
then $m_p$ eventually becomes larger than $M_{d0}$, such that 
Eq.~(\ref{rdot-Ivanov}) may apply. 
Thus, the parameter regime that we consider in 
this paper is expected to be transitional between those
governed by  Eqs.~(\ref{rdot-II})
and~(\ref{rdot-Ivanov}).

\section{Protostellar Disc Models} \label{discs}
Models of protostellar discs considered
as viscous accretion discs have been constructed by a number
of authors (e.g. Bell {\em et al}. 1997, Papaloizou, Terquem, \& Nelson 1999).
An important issue is
the nature of the  effective viscosity. Usually  the `$\alpha$'
 prescription of Shakura \& Sunyaev (1973),  
$\nu = \alpha H^2 \Omega$ is adopted. The most likely
 source of  the turbulence required to produce the effective
viscosity is MHD instabilities (Balbus \& Hawley 1991, 1998). 
Simulations have shown that values of $\alpha$ in the range
$10^{-3} - 10^{-2}$ may be produced. However,
it is unclear that adequate ionization exists for the mechanism 
to work throughout the disc and  in some cases MHD instabilities
may exist only in a surface layer ionized by cosmic rays (Gammie 1996).

Assuming MHD instabilities do work throughout
the disc and  produce values of $\alpha$ in the above range, disc models
with properties similar to that of a minimum mass solar nebula
are produced at a time $\sim 10^{5}$ -- $10^{6}$~yr after formation
(Papaloizou, Terquem, \& Nelson  1999).
These typically have $H/r \sim 0.04 - 0.05$ so that the kinematic viscosity
$\nu \sim 10^{-5}$ at 5~AU.

Although there are uncertainties as to how a solid core of sufficient
mass accumulates for rapid gas accretion to begin, in order to 
understand the distribution of the  orbital parameters of extrasolar
planets,
it is reasonable
to pose the question as to how a Jupiter mass protoplanet evolves
as a result of interaction with the protostellar disc immediately
post formation.
 
Accordingly we shall consider the interaction of a Jupiter mass
protoplanet with a disc 
containing two Jupiter masses interior to the initial protoplanet  orbit
and with a constant $\nu = 10^{-5}.$
We consider cases where the initial disc surface density is uniform
and where it  depends on radius.

\section{The Physical Model} \label{phys-mod}
\subsection{Equations of Motion} \label{equations}
The vertical thickness $H$ of an accretion disc, which is in a state
of near--Keplerian rotation, is small in comparison to the distance $r$
from the centre, {\em i.e.} $H/r \ll 1$. It is therefore convenient to
vertically average the equations of motion and work with vertically averaged
quantities only, the assumption being that there is zero vertical motion.
 The problem  is thus reduced to a 2--dimensional one.

We shall work with cylindrical coordinates ($r$, $\phi$, $z$), with
the origin located at the position of the central star. 
The velocity in our 2--dimensional disc is denoted by 
${\bf v}=(v_r$, $v_{\phi}$, 0), where $v_r$ is  the radial velocity
and $v_{\phi}$ is the azimuthal velocity. We denote the angular
velocity of the disc material by $\Omega=v_{\phi}/r$, where the rotation axis
is assumed to be coincident with the vertical axis of the coordinate system.
The vertically integrated
continuity equation is given by
\be \frac{\partial \Sigma}{\partial t} + \nabla . (\Sigma {\bf v}) =0.
\label{continuity} \ee
The components of the momentum equation are
\be \frac{\partial (\Sigma v_r)}{\partial t} + \nabla . (\Sigma v_r {\bf v})
= 
\frac{\Sigma v_{\phi}^2}{r} 
- \frac{\partial P}{\partial r} - 
\Sigma \frac{\partial \Phi}{\partial r} + f_r \label{r-momentum} \ee
\be \frac{\partial (\Sigma v_{\phi})}{\partial t} +
\nabla . (\Sigma v_{\phi} {\bf v}) =
- \frac{\Sigma v_rv_{\phi}}{r} 
-\frac{1}{r} \frac{\partial P}{\partial \phi} 
- \frac{\Sigma}{r} \frac{\partial \Phi}{\partial \phi} + f_{\phi}
\label{phi-momentum} \ee
Here $\Sigma$ denotes the surface density
$$ \Sigma = \int^{\infty}_{-\infty} \rho dz, $$
with $\rho$  being the density, $P$ is the vertically integrated pressure, 
and $f_r$ and 
$f_{\phi}$ denote the viscous force per unit area acting in the
$r$ and $\phi$ directions respectively.
The gravitational potential, $\Phi$, is given by
\begin{eqnarray} 
\Phi = & - & \frac{G M_*}{r} 
- \frac{G m_p}{\sqrt{r^2 + r_p^2 - 2 r r_p \cos{(\phi - \phi_p)}}} \nonumber \\
& + & \frac{G m_p}{r_p^3} {\bf r}.{\bf r}_p 
+ G \int_S \frac{dm({\bf r}')}{r'^3} {\bf r}.{\bf r}' \label{potential} 
\end{eqnarray}
where $M_*$ and $m_p$ are the masses of the central star and the protoplanet
respectively, and $r_p$ and $\phi_p$ are the radial and azimuthal coordinates
of the protoplanet. The third and fourth terms in Eq.~(\ref{potential})
account for the fact that the coordinate system based on the
central star is accelerated by the combined effects of the protoplanetary
companion and by the gravitational force due to the disc, respectively, with
the integral
in Eq.~(\ref{potential}) being performed over the
surface of the disc.

In our models the protoplanet evolves under the  gravitational attraction of
the central star and 
 the protostellar disc. The latter  interaction is expected to cause
the protoplanet to undergo orbital evolution. The 
equation of motion  of the protoplanet may be written
\be \frac{d^2{\bf r}_p}{dt^2} = -\frac{G (M_* + m_p)}{r_p^3}{\bf r}_p -
\nabla \Phi_d \label{planet-accel}  \ee
where the gravitational potential of the disc is given by
\be \Phi_d = -G \int_S \frac{\Sigma({\bf r}')}{|{\bf r}' - {\bf r}_p|} 
d{\bf r}'
+ G \int_S \frac{dm({\bf r}')}{r'^3} {\bf r}.{\bf r}'.
\label{disc-pot} \ee
Here the integrals are performed over the disc surface, and the second term
is the indirect term arising from the fact that the coordinate system is
accelerated  by the disc gravity ---~note that the part of the indirect
term due to the planet itself is 
already included in Eq~(\ref{planet-accel}). As the 
protoplanet orbits about
the central star, it is able to accrete gas from the surrounding 
protostellar disc. The disc gas that it accretes has an associated specific
angular momentum, which if different from the specific angular momentum
of the protoplanet, will cause its orbit to evolve. We include the effects
of this `advected angular momentum' on the protoplanet's orbit
by calculating the appropriate
changes to the planet's mass and angular momentum that arise from
the gas accretion at each time step.

\subsection{Equation of State}
For computational simplicity we adopt a locally 
isothermal equation of state.
The vertically integrated pressure is related to the surface density through
the expression 
\be P = c_s^2 \Sigma \label{E-state} \ee
where the local isothermal sound speed is  given by
$$c_s = \frac{H}{r} v_K$$
where $v_K=\sqrt{G M_* / r}$ denotes the Keplerian velocity of the
unperturbed disc. The disc aspect ratio, $H/r$, is assumed to be an input 
parameter that defines the Mach number of the disc model being considered.
The calculations presented in this paper  for the most part
adopted $H/r=0.04.$
 Some calculations  denoted by R{\it i}  
adopted $H/r=0.05$, see table~\ref{tab1}.

\subsection{Viscosity}
 In this present work, we assume that protostellar discs
have an anomalous effective viscosity  most probably arising from
MHD turbulence (see discussion in section \ref{discs}).
 Here we make the assumption
that  this effective viscosity
can be modeled by simply replacing the molecular
kinematic viscosity coefficient in the Navier--Stokes equations
by a turbulent viscosity coefficient denoted by $\nu$. 

The components of the viscous force  per unit area may  then be written
\be f_r = \frac{1}{r} \frac{\partial (r \tau_{rr})}{\partial r}  +
 \frac{1}{r} \frac{\partial \tau_{r \phi}}{\partial \phi} - 
 \frac{\tau_{\phi \phi}}{r} \label{visc-r} \ee
 \be f_{\phi} = \frac{1}{r} \frac{\partial (r \tau_{\phi r})}{\partial r} +
 \frac{1}{r} \frac{\partial \tau_{\phi \phi}}{\partial \phi} + 
 \frac{\tau_{r \phi}}{r}, \label{visc-phi} \ee
where the components of the viscous stress tensor used in the above
expressions are
\begin{eqnarray}
\tau_{rr} & = & 2 \eta D_{rr} - \frac{2}{3} \eta \nabla . {\bf v} \\ \nonumber
\tau_{\phi \phi} & = & 2 \eta D_{\phi \phi} - \frac{2}{3} \eta \nabla . 
{\bf v}
\\ \nonumber
\tau_{r \phi} & = & \tau_{\phi r} = 2 \eta D_{r \phi}, \label{visc-tensor} 
\end{eqnarray}
where 
\begin{eqnarray}
D_{rr} &=& \frac{\partial v_r}{\partial r}, D_{\phi \phi} = \frac{1}{r}
\frac{\partial v_{\phi}}{\partial \phi} + \frac{v_r}{r} \\ \nonumber
D_{r \phi} &=& \frac{1}{2} \left[ r \frac{\partial}{\partial r} \left(
\frac{v_{\phi}}{r} \right) + \frac{1}{r} \frac{\partial v_r}{\partial \phi}
\right], \label{D} 
\end{eqnarray}
and $\eta=\Sigma \nu$ is the vertically integrated
 dynamical viscosity coefficient.
In the work presented in this paper we use a constant value for
$\nu=10^{-5}$.

\subsection{Dimensionless Units} \label{Units}
For reasons of computational simplicity, we use dimensionless units for
our numerical calculations. The unit of mass is taken to be the sum of
mass of the central star $M_*$ and the 
initial mass of the protoplanet $m_p$,
and the unit of length is taken to be the 
initial orbital radius of the protoplanet $a_0$.
We set the gravitational constant $G=1$.
The unit of time then becomes 
$$ t_0 = \sqrt{\frac{a_0^3}{G(M_*+m_p)}}. $$
When discussing the results of the calculations in subsequent sections,
we will use the initial orbital period of the planet as the unit of
time, given by $P_0 = 2\pi t_0$.

\subsection{Gas Accretion by the Protoplanet} \label{accretion}
Following Kley (1999a), the accretion of gas by the protoplanet 
is modeled by removing 
a fraction of the material that resides within a distance of
$\frac{1}{2}R_r$ from the protoplanet at each time step, and a different
fraction from within $\frac{1}{4}R_r$, where
$R_r$ is the Roche radius of the planet given by
\be R_r = r_p \left(\frac{m_p}{3M_*} \right)^{1/3} \label{Roche} \ee
The fraction of gas removed at each time step determines the
local accretion time scale onto the protoplanet, $t_{acc}$,
which in our calculations is
taken to be $t_{acc} = 3t_0$ within $\frac{1}{2}R_r$ and $t_{acc}=1.5 t_0$
within $\frac{1}{4}R_r.$
This accretion rate is large
and is  almost  maximal (Kley 1999a). We also
perform simulations in which the accretion rate is set to zero.
In this case a lobe filling atmosphere develops such that
material that approaches the protoplanet is  forced by its pressure
to  either return or cross the gap.  Thus
 our calculations span the range of possible
accretion rate behaviour onto the protoplanet. 

\subsection{Initial  Conditions}
\label{ini-bc}
The disc models used for all simulations using the codes NIRVANA and
FARGO  
were of uniform surface density initially, 
 had a value of $\nu=10^{-5}$ throughout and a constant value of $H/r=0.04$.
The value of $\Sigma$ was chosen such that there exists the equivalent of
2 Jupiter masses in the disc interior to the orbital radius of the protoplanet,
initially. 
 In our dimensionless units this gives
$\Sigma=6 \times 10^{-4}$. 
The initial mass ratio between the protoplanet and the 
central star was taken to be $q=m_p/M_*=10^{-3},$ corresponding to a
Jupiter mass planet orbiting about a solar mass star. 
The protoplanet  was started on a circular orbit of radius $r=1.$ 
The inner radius of the disc is located at $r=0.4$ and the outer radius is
located at $r=2.5$.

Simulations with RH2D  were carried out using identical initial conditions
to those described in Kley (1999a). The aspect ratio
of the disc was  $H/r=0.05.$
The initial 
density profile was  of the form
$\Sigma(r) \propto r^{-1/2}$, and in all cases
an initial annular gap was imposed. The inner  disc radius is
located at $r=0.25$ and the outer radius
is at $r=4.0.$. The total initial disc  mass
was $10^{-2} M_{\odot}.$  In comparison, the initial models for NIRVANA
and FARGO have  a mass $1.45$ times larger. Hence, 
the surface density of the models with RH2D
at $r=1$ is about three times
lower.                        

\subsection{Boundary Conditions}
An outflow boundary condition is used at the inner 
boundary for all calculations presented here since material in a viscous 
accretion disc will naturally flow onto the
central star. 

The outer boundary condition is more problematic, since ideally we
would like to have a closed outer boundary.
We work in a coordinate system that is based on the central star and not
on the centre of mass. The natural tendency for material at the outer edge of 
the disc is to orbit about the centre of mass and not the central star.
Adopting the usual closed boundary
condition of $v_r=0$ and $v_{\phi}=\sqrt{G M_*/r}$ will
result in a mismatch between the disc material just interior to the outer
boundary and that imposed at the boundary itself. Calculations using this
condition do indeed show the resulting excitation of waves at the 
outer edge of the disc,
even for the case of a non--accreting planet on a fixed circular orbit.

In order to alleviate this problem, the boundary condition that we adopt
in the NIRVANA calculations assumes that material at the
outer edge of the disc is in circular, 
Keplerian orbit about the centre of mass
of the star plus planet system. Since we work in a coordinate system based 
on the central star, this requires us to calculate the correct values of
$v_r$ and $v_{\phi}$ at each time step and apply them to the outer
boundary. One consequence of this is that the outer boundary is no longer 
closed, but can allow the inflow and outflow of material since
the radial velocity is no longer zero. This effect is small, however, and 
appears to have a negligible effect on our simulation results.
Simulations of a non--accreting planet
on a fixed circular orbit using this boundary condition show no signs of
wave excitation at the outer edge of the disc. When the planet is able
to accrete gas from the disc and to orbitally migrate, this boundary condition
again results in some wave excitation at the outer edge, but with a much
reduced amplitude. This arises because the centre of mass position
changes with time as the planet grows in mass and changes its orbit,
and these changes are fed into the outer boundary condition
instantaneously. The disc material interior to the outer boundary
on the other hand, will adjust to the centre of mass evolution
on a longer time scale, so that this boundary condition also produces
a small but noticeable mismatch between the boundary and the outer disc
material.

The calculations performed using FARGO and RH2D all used a closed
outer boundary
condition such that $v_r=0$. The strong similarities in the results
of the calculations performed with NIRVANA and FARGO indicate
that the details of the outer boundary have a negligible effect on our results.
This is because, as tests have shown,  the
protoplanet primarily interacts with material that is close to its
immediate vicinity.

\section{The Hydrodynamic codes} \label{codes}
In order to establish the  reliability
of the numerical results, 
the equations of motion~(\ref{continuity}) to~(\ref{phi-momentum}) have been
solved using three different Eulerian hydrodynamic codes,
NIRVANA, FARGO and RH2D.
In each case the equations are solved using a finite difference scheme on a 
discretised computational domain  containing  $N_r \times N_{\phi}$
grid cells. Each  scheme is described briefly below.

\begin{figure*} 
\epsfig{file=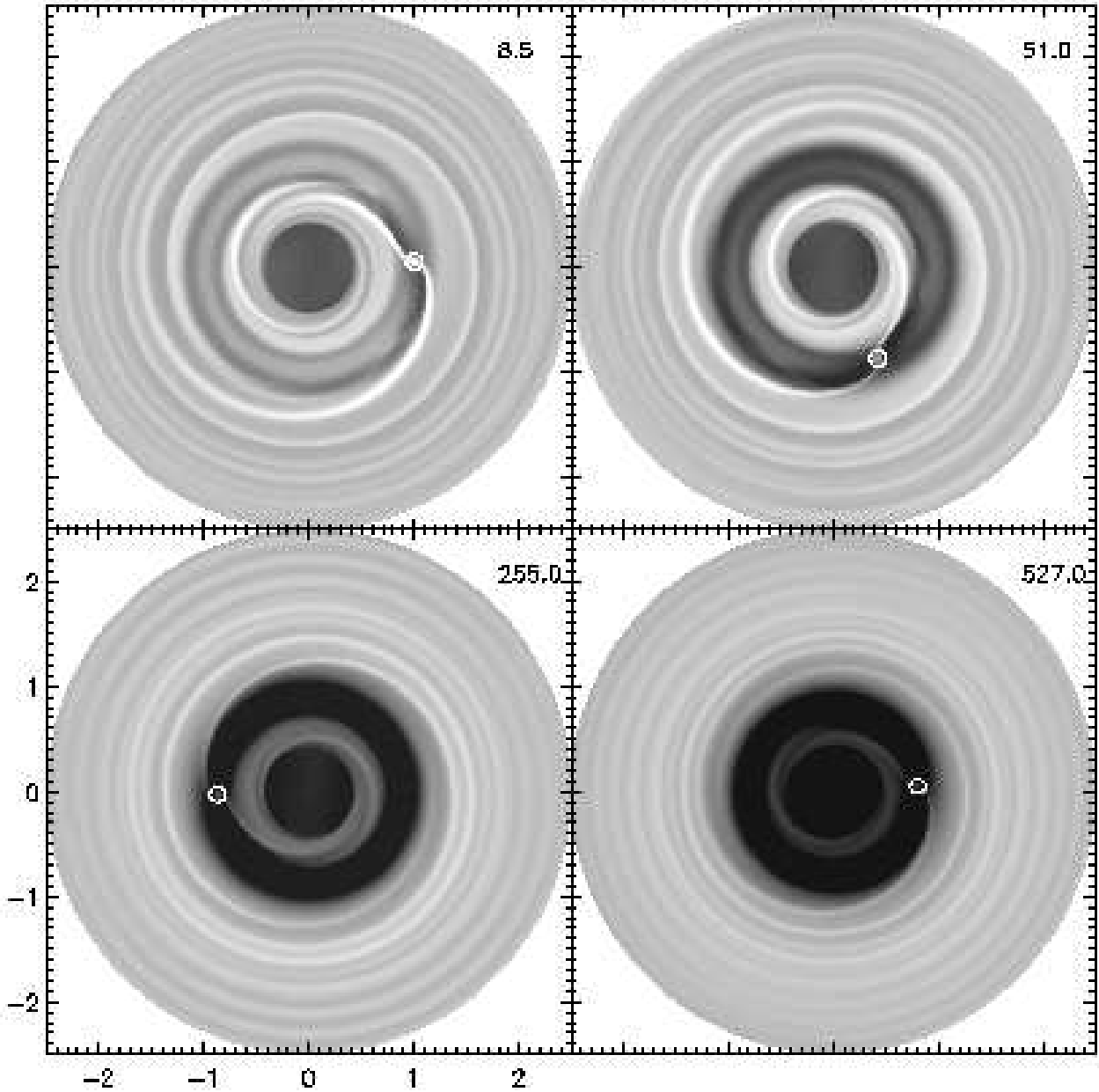, width=\textwidth}
\caption{This figure shows the evolution of a protoplanet embedded
in a protostellar disc for calculation N3.
The relative surface density of disc material is
represented by the grey-scale. The white circle represents the position
of the protoplanet. The disc is initially unperturbed at time $t=0$.}
\label{rpn1}
\end{figure*}

\subsection{NIRVANA}
NIRVANA is a three dimensional MHD code that 
has
been described in depth elsewhere (Ziegler \& Yorke 1997).
For the simulations presented here, the magnetic field is set to zero
 such that the code   becomes purely hydrodynamic.  We work
in two dimensions and  use cylindrical
 ($r$, $\phi$)
coordinates. Viscous terms  have been added as described
by Kley (1998). The computational domain is subdivided into $N_r \times
N_{\phi}$ zones, where the grid spacing in both coordinate directions is uniform.

For the calculations that are
presented in this paper, three different levels of resolution have been used.
The low resolution runs use $N_r=50$ and $N_{\phi}=147$, the
mid--resolution runs use $N_r=80$ and $N_{\phi}=235$, and the
high resolution runs use $N_r=130$ and $N_{\phi}=384$.
The numerical method is based on a spatially second--order accurate, explicit
method that computes the advection using the second order 
monotonic transport algorithm
(Van Leer 1977), leading to the global conservation of mass and angular 
momentum. 
The evolution of the planet orbit is computed using a standard leapfrog
integrator.
NIRVANA has been applied to a number of different problems  including
that of an accreting protoplanet embedded in a protostellar disc.
It was found to give results that are very similar to those obtained with 
other finite difference codes including  RH2D (e.g. Kley 1999a).
\subsection{FARGO}
 This is an alternative Eulerian ZEUS-like code, based on the FARGO
fast advection method (Masset 1999). The main difference between
 NIRVANA  and this code is that in  FARGO
 the time step is not limited by the classical CFL condition,
which results in a very small time step due to the fast orbital motion
at the inner boundary, but it is limited by a CFL condition based on the residual
velocity with respect to the average orbital motion. This leads to a
substantially larger time step and hence faster computation. Also, 
the FARGO procedure leads to a smaller numerical diffusivity because a larger
time step size requires one to perform fewer advection substeps during the
calculations. 
Since the time step in the FARGO simulations can be quite large, 
especially in the low resolution case, 
a fourth-order
Runge Kutta scheme was used to integrate the equations
of motion of the protoplanet.

Since this work represents the first  application of the FARGO
advection algorithm, it has been widely tested against NIRVANA by
running strictly similar simulations (identical physical and numerical
parameters). The good agreement between both codes, along
with the low numerical diffusivity of FARGO, have validated
FARGO as a very useful tool  for studying
 in the embedded protoplanet problem.
Since it is
much faster than NIRVANA, it has been used to  calculate the very 
long-term
behaviour of an accreting protoplanet (see run F6 below).
\subsection{RH2D}
\label{code:rh2d}
 To obtain results using a 
 third method we  have employed the code RH2D (Kley 1989).
This has been used previously
in studies of disc-protoplanet interaction (Kley 1998, 1999a).
It is a two-dimensional radiation hydrodynamics code. 
For the  simulations presented here
the radiation module is switched off, and all parts are solved
explicitly.
The code is based on the second order Van Leer (1977) advection algorithm
and uses a staggered grid, with logarithmic spacing
(a constant enlargement ratio of neighbouring grid cells)  
in radius and which is uniform in 
 azimuth. 
As in FARGO, a fourth order Runge Kutta
scheme was used to integrate the protoplanet orbit.
 This code has already been found to give similar results to NIRVANA
for protoplanet problems (Kley 1998).
Accordingly we here use  RH2D to study the orbital
evolution of an embedded planet under slightly different conditions
 to those adopted in the case of the other two numerical methods. 

\section{Numerical Calculations} \label{calculations}
The main results of the numerical calculations are presented in 
table~\ref{tab1}. Of particular  significance 
are the migration times, $\tau_{mig}$, which are listed in the fifth
column of table~\ref{tab1} in units of $10^4 P_0$, and the estimated final
masses of the planets, $m_{final}$,
which are listed in the sixth column in units
of the original planet mass $m_{init}.$

The values of $\tau_{mig}$ were obtained by measuring the
rate of change of the planet's semi-major axis at the end of the simulation
and extrapolating forward in time assuming that this rate remains constant.
Thus $\tau_{mig}$ gives as estimate of the time required for the planet to
migrate all the way  to the central star. The values of $m_{final}$ 
were obtained
by measuring the accretion rate at the end of the simulation and 
extrapolating forward for a time $\tau_{mig}$.

Thus $m_{final}$ provides an
estimate of the  mass the planet will attain
on  migrating all the way  to the
central star. It should be noted that while these values provide 
reasonable estimates of the true migration times and accreted masses, the 
assumption of there being a constant migration or accretion rate 
is not strictly  correct since these will tend to decrease as the planet
mass continues to grow and migrate inwards.
The only difference between runs N$i$ and F$i$ ($i \in [1,5]$) is
the code used to perform the calculations, with the N$i$ runs being performed
with NIRVANA and the F$i$ runs with FARGO. The evolution equations of
the system and the initial  conditions are identical. Their results agree
reasonably well, except for the low resolution non accreting case,
but as we shall see below, the low resolution we used for these runs
is probably too coarse to give trustworthy results. From both the mid-resolution
(N3 \& F3) and high resolution (N5 \& F5) runs, we can see that  FARGO
 gives slightly higher accretion rates (and consistently
longer migration times, since the inertia of the planet increases 
and, less importantly, most of the accreted material
comes from the outer disk with a larger specific angular momentum).
This higher accretion rate arises
because  FARGO  has a smaller numerical diffusivity
along the direction of orbital motion (Masset 1999).
Associated with this is a more strongly peaked
density profile 
around the planet.
\begin{table}
\begin{center}
\begin{tabular}{lllllr}  \hline \hline
 Run &  Resolution & Accretion   & Initial & $\tau_{mig}$ & $m_{final}$ \\
     & $N_r \times N_{\phi}$ & on ? & Gap ? & $\times 10^4 P_0$ & ($m_{init}$) \\ \hline
   &    &             &        & & \\
   N1 & $50 \times 147$ & Yes & No & 1.77 & 4.8 \\
   N2 & $50 \times 147$ & No & No & 1.07 & 1.0 \\
   N3 & $80 \times 235$ & Yes & No & 1.00 & 3.2 \\
   N4 & $80 \times 235$ & No & No & 1.20 & 1.0 \\
   N5 & $130 \times 384$ & Yes & No & 0.84 & 2.7 \\
   N6 & $130 \times 384$ & No & No & 1.0 & 1.0 \\
   N7 & $80 \times 235$ & Yes & Yes & 0.45 & 3.25 \\
   N8 & $80 \times 235$ & No & Yes & 0.40 & 1.0 \\
\hline
   F1 & $50 \times 148$ & Yes & No & 1.99 & 4.77 \\
   F2 & $50 \times 148$ & No & No & 1.75 & 1.0 \\
   F3 & $80 \times 236$ & Yes & No & 1.08 & 3.85 \\
   F4 & $80 \times 236$ & No & No & 1.33 & 1.0 \\
   F5 & $130 \times 384$ & Yes & No & 0.85 & 3.27 \\
   F6 & $70 \times 180$ & Yes & Yes & 2.03 & 4.87 \\
\hline
   R1 & $128 \times 128$ & (No) & Gap & 1.45 & 1.0 \\
   R2 & $128 \times 128$ & Yes & Gap & 1.48 & 4.63 \\
   R3 & $128 \times 288$ & Yes & Gap & 1.36 & 4.10 \\
\end{tabular}
\end{center}
\caption{ \label{tab1}
The first column gives the run label (N$i$ NIRVANA, F$i$ FARGO,
R$i$ RH2D), the second column gives
the number of grid cells used, and the third column indicates whether
the planet accretes from the disc. The fourth column indicates whether
the calculation started with an initial gap in the disc. The fifth column
gives the estimated migration time (in units of $10^4 P_0$) and the
sixth column provides an estimate of the final mass of the planet.
Note that the estimated times are  given  at different evolutionary times
of the models.
}
\end{table}

For the purposes of illustrating the main results of our simulations,
we will now describe the results for one individual case. Following this
we will compare the results of simulations in which the disc either did or
did not have a tidally induced gap in  which the planet orbits at the start
of the calculation. We will then go on to compare the 
migration rate of 
a planet  which accretes gas from the disc at a maximal rate to that of
a non accreting protoplanet.
We then look at the effect of changing the initial surface density profile
and disc aspect ratio.
Following this, we compare the results obtained with
NIRVANA and FARGO , and study  the effects of changing the numerical
resolution.
\subsection{An Illustrative Case \label{standard-run}} 
\begin{figure} 
\epsfig{file=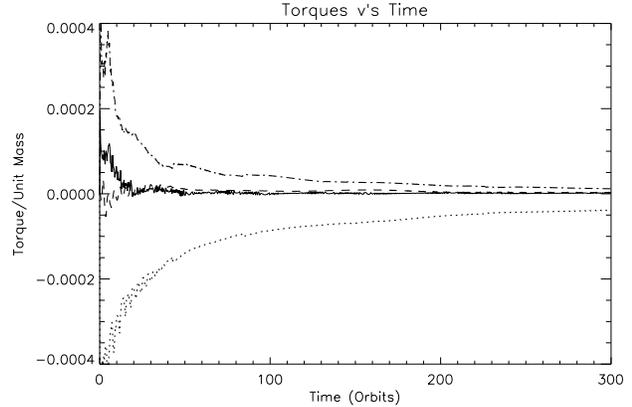, width=8.5cm}
\caption{\label{rpn2} This figure shows the torques 
(per unit mass) acting on the
protoplanet during the early stages of its evolution, in run~N3. 
The torque due to the
disc exterior to the planet orbital radius
$r_p$ is shown by the dotted line, that due to the
disc interior to or at $r_p$ is shown by the dot-dashed line, and the torque
due to the disc's contribution to the indirect term of the gravitational
potential is shown by the dashed line. The effective torque due to
the accretion of disc gas is shown by the solid line.}
\end{figure}

The evolution of the protoplanet embedded in the disc for calculation N3
is shown in fig.~\ref{rpn1},
where grey-scale plots
showing the surface density variation in the disc are presented,
and the position of the planet is indicated by the small white circle
(which has a radius equal to the Roche radius of the planet).
Initially at time $t=0$
the disc
surface density was unperturbed, but as the calculation proceeds 
the tidal force due to the planet strongly perturbs the disc,
leading to the formation of trailing spiral shock waves. In particular
an $m=2$ spiral wave pattern may be observed. The transfer of 
angular momentum between the disc and the protoplanet leads to the formation
of an annular gap, or surface density depression, in the vicinity of
the planet's orbit, which is cleared after about 200 orbits for a
Jupiter mass planet. 

As the disc--planet system evolves, the inner disc is lost from the system
since viscous evolution causes it to
  drain through the open inner boundary. The tidal interaction
between the disc and protoplanet leads to the disc interior to $r_p$ 
exerting positive torques on the protoplanet, and the disc exterior
to $r_p$ exerting negative torques on it.
The loss of the inner disc leads to a reduction of the positive 
torque, so that the torque due to the outer disc becomes 
dominant. Consequently the planet undergoes inward orbital migration as
its angular momentum is removed by the outer disc. In the final panel of
fig.~\ref{rpn1}, which corresponds to a time of $t=527 \;\; P_0$,
it may be observed that the semi-major axis of the planet's
orbit is $\sim 80$ percent of it's original value, and that most of the
inner disc has  disappeared. As the calculation proceeds beyond this
point the planet continues to spiral in towards the central star.

The evolution of the torques acting on the protoplanet during the first
300 orbits are shown in 
fig.~\ref{rpn2}, and indicate what the dominant contribution 
to the orbital migration is.
Here, the torque per unit mass acting on the
protoplanet due to the disc exterior to the orbital radius of the
protoplanet $r_p$ (dotted line),
interior to, or at, $r_p$ (dot-dashed line), and the indirect term in the
gravitational force (dashed line) are plotted against time.
Also plotted is the effective torque per unit mass that arises from the
accretion of material whose specific angular momentum differs from
that of the orbit (solid line). 
The first thing to note is that the accretion of gas
from the disc has a negligible effect on the orbital evolution. 
 Once the gap has formed the effective torque
arising from  accretion is only $\sim 3$ percent of the torque  due to
the outer disc. Second, we note that the torque arising from the 
outer disc material is consistently larger than that due to the
inner disc material. The loss of the inner disc through the open 
inner boundary causes this disparity in the torques to grow, and ensures 
that the migration is always directed inwards. The contribution that the
disc gravity makes to the indirect term is found to have a
negligible effect on the orbital evolution.
\begin{figure*} 
\epsfig{file=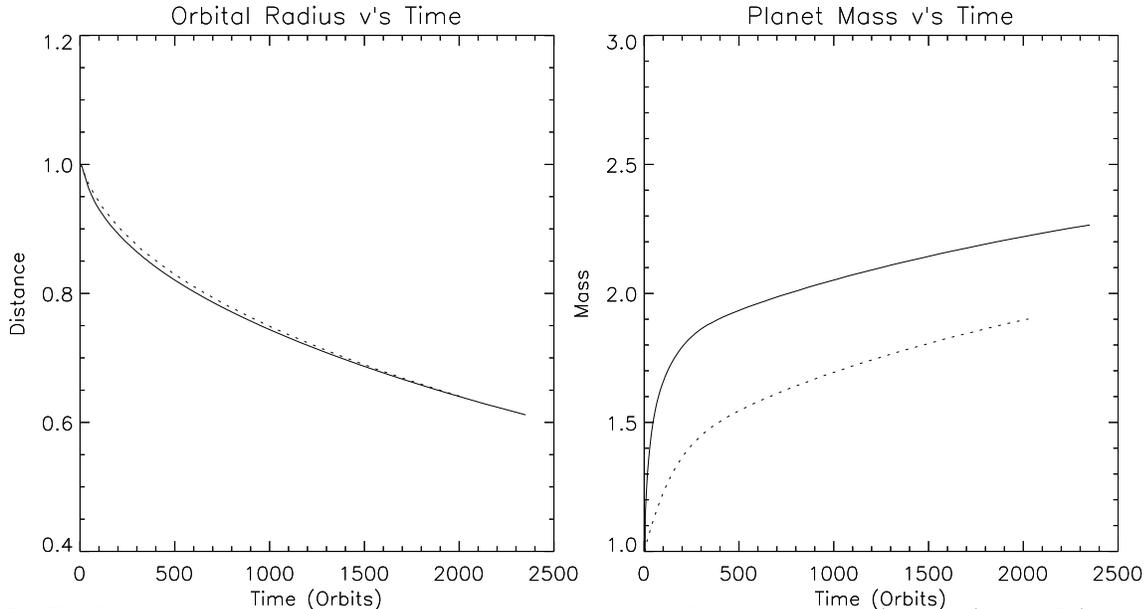, width=15cm}
\caption{\label{rpn3} The first panel shows the evolution of the 
planet orbital radius for calculation N3 (solid line) and N7 (dotted line).
The second panel shows the increase in mass as the protoplanets accrete
gas from the disc for the same calculations, with the same line styles 
as described above.}
\end{figure*}

The evolution of the orbital radius of the protoplanet is shown by the
solid line in the first panel of fig.~\ref{rpn3}, and that of the
planet mass is shown by the solid line in the second panel.
The evolution 
of the protoplanet undergoes a more rapid phase early on as the
planet loses angular momentum to the disc material during the gap opening phase,
leading to a faster period of orbital migration.
This phase of evolution lasts for a few hundred orbits until the gap is
opened, and also results in a very large mass accretion rate that almost doubles
the mass of the protoplanet within $t=300\;\;P_0$.
Once the gap has been cleared, the evolution slows down and the planet
spirals in towards the star at an almost constant rate while accreting gas
from the disc at an almost steady accretion rate. The migration rate is
observed to slow down 
slightly as the calculation proceeds since the increase
in mass of the planet increases its inertia. 
The calculation was initiated with $M_{d0} \sim 2 m_p$, but as the planet 
migrates and accretes from the disc we arrive at a situation 
where $m_p \ge M_{d0}$ (see discussion in section ~\ref{typeII}).
The orbital evolution of
a massive planet orbiting in a disc that contains a smaller mass than the 
planet within a characteristic radius $r_p$ has been studied
by Ivanov, Papaloizou, \& Polnarev (1999). In this physical regime the evolution
is controlled by the viscous evolution of the disc and the inertia
of the planet, 
with a more massive planet migrating more slowly, as described
in section~\ref{typeII}.
We can  test whether  Eq.~(\ref{rdot-Ivanov}) agrees
with our numerical calculations by examining the changes 
in planet mass, orbital 
radius, migration rate, and disc surface density at two different times,
$t_1$ and $t_2$, during the calculation. If we allow for the reduction
of the surface density 
in the outer disc due to accretion onto the planet, and by mass 
flow through the gap and  through the inner boundary, then we obtain the
following expression relating the migration rates ${\dot r}_p(t)$
\be \frac{{\dot r}_p(t_1)}{{\dot r}_p(t_2)} = 
\left( \frac{m_p(t_2) \Sigma(t_1)}{m_p(t_1) \Sigma(t_2)} \right)^{4/5}
\left( \frac{r_p(t_1)}{r_p(t_2)} \right)^{3/5} \label{rdot-t1t2} \ee
for a disc with a constant value of $\nu$. Comparing the migration rates
in the numerical
calculation at $t_1=1000 \; P_0$ and $t_2=2000 \; P_0$, we find reasonable
agreement between the prediction of   Eq.~(\ref{rdot-t1t2})
and the numerical results. The measured ratio of the migration rates
gives ${\dot r}_p(t_1)/{\dot r}_p(t_2) \simeq 1.5$, whereas that predicted 
by Eq.~(\ref{rdot-t1t2}) gives a value 
${\dot r}_p(t_1)/{\dot r}_p(t_2) \simeq 1.4$ with $\Sigma$ 
taken  as the azimuthally averaged value at the location
of the 2:1 outer Lindblad resonance.

The accretion rate
 decreases as the protoplanet  mass increases because the disc is tidally
truncated more effectively by a more massive planet (e.g. BCLNP)
After a time of
$t \sim 2400 \;\; P_0$, the protoplanet has migrated to a radius of
$r \sim 0.6$ and has accreted $\sim 1.3$ times its original mass.
By extrapolating the migration and accretion rates forward in time,
it is estimated that the planet will spiral into the central star after
$t \sim 1.0 \times 10^4 \;\;P_0$, by which time it will
have reached a mass 
$m_{final} \sim 3.2$ times its original mass. 

We note that the planet remains
in an almost circular orbit throughout its evolution, and shows no
sign of eccentricity growth, since the gap region still contains sufficient
corotating material that is able to damp the eccentricity growth
caused by the outer Lindblad resonances 
(Goldreich \& Tremaine 1980, 
Artymowicz 1993a,b).

The viscous evolution
time of the disc is given by Eq.~(\ref{visc-time}).
At a radius of $r=1$, this corresponds
to an evolution time of $\sim 10^4 \;\; P_0$, very similar to what is
observed for the migration time of the protoplanets in all of our simulations.

This confirms the idea described in section~\ref{typeII} 
and in Lin \& Papaloizou
(1986) that giant protoplanets undergoing tidal interaction
with a protostellar disc should  migrate on a time controlled
by the viscous evolution time of the disc when the 
interaction is sufficiently
non linear to open up a gap, and when the mass of the planet is less
than  or comparable
to the disc mass with which it gravitationally interacts.


\subsection{The Effects of an Initial Gap} \label{gap-no-gap}


As well as 
performing calculations in which the initial disc was unperturbed,
we also performed calculations in which the initial disc contained 
a tidally induced
gap around the vicinity of the planet. Here we will concentrate on the 
calculations labelled as N3 and N7 in table ~\ref{tab1}.

The calculation labelled as N3 is for an accreting protoplanet
embedded in an initially unperturbed disc, and was described  in detail
in the previous section~\ref{standard-run}.
The calculation labelled as N7 in table~\ref{tab1} is for an accreting
protoplanet initially embedded in a disc which has a tidally induced
gap at time $t=0$. This initial condition was obtained by running
a calculation with a non accreting
planet on a fixed circular orbit for $t \sim 300 \; P_0$,
until a 
clear gap was formed and the surface density in the gap region became
steady. 
A reflecting inner boundary condition was employed during this phase in
order that the disc mass was conserved. The results of this
calculation were then used as the initial conditions of calculation
N7, but now with an accreting protoplanet which was able to undergo
orbital evolution, and with an open inner boundary condition. 

The evolution of the orbital radius of the planet in calculation N7
is shown by the
dotted line in the first panel of fig.~\ref{rpn3}, and the evolution of
the mass of the planet is shown by the dotted line in the second panel.
In the case of the calculation N3, in which the planet is initially
embedded in an unperturbed disc and must clear material in order to
form a gap, the clearing of that material leads to a period of more rapid
migration. At the same time, this larger migration rate is slowed
by the
rapid accretion of gas, when the planet is deeply embedded, 
that leads almost to a doubling of the
planet mass and inertia within a few hundred orbits. 
In the case of calculation N7, the planet initially resides within a gap,
and so does not have to clear much material away from its vicinity 
during the early stages of its orbital evolution. 
It does, however, have an early period of more
rapid accretion since it absorbs the material that is initially within
its Roche lobe that accumulated there during the formation of the gap.
This initially large accretion rate is augmented by the fact that
an accreting planet helps to reduce the surface density
of material in the gap region by accreting some of it.
The similarity of the migration rates for the calculations N3 and N7
during the first $1000 \; P_0$ indicates that the effects of opening the gap
for an initially fully embedded planet are almost entirely
counter balanced by the planet mass growth and accretion of material
with a higher specific angular momentum, with this latter effect being 
almost negligible after $\sim 100 \; P_0$ (as shown in fig.~\ref{rpn2}).

Looking at the the later stages of the evolution in the first panel of 
fig.~\ref{rpn3}, 
we notice that although the planet in calculation N3 migrates
slightly faster to begin with, 
the planet in calculation
N7 eventually migrates at a higher rate since its mass and inertia 
are smaller, with the orbital radii crossing over at $t \sim 1500 \; P_0$.
The expected migration time scale in calculation N7 is 
$\tau_{mig} \sim  0.45\times 10^4 \;P_0$, with the final mass estimate being 
$m_{final} \sim 3.2$, indicating that the presence
or not of an initial gap has only a relatively small effect on the final
results.

\subsection{Comparison of the Evolution of an Accreting  
and Non Accreting Protoplanet} \label{accret-no-accret}

Calculations were performed for both accreting and non accreting planets.
In this section we will compare the results of calculations N3 and N4 in
order to ascertain the effects of  accretion  on the evolution of the
protoplanet.  These calculations were both initiated with
unperturbed disc models.
For the accreting protoplanets, the accretion rate adopted  is 
such that the e--folding time for the protoplanet to accrete  mass 
within  a distance of half
of its Roche radius is given by $\tau_{acc}=3 P_0/(2 \pi).$ This is a 
dynamical  time scale  so the
situation corresponds essentially to a maximally accreting protoplanet
(Kley 1999a).

The case of a non accreting planet  corresponds to the situation 
where the protoplanet has a lobe filling gaseous
envelope in hydrostatic equilibrium with 
 Kelvin--Helmholtz time  longer
than the migration time. Then the planet can accrete  little mass during
the migration stage. Such a situation
may  occur if the envelope  is built up while
maintaining a low luminosity emitted  from the central parts of the protoplanet.

\begin{figure} 
\epsfig{file=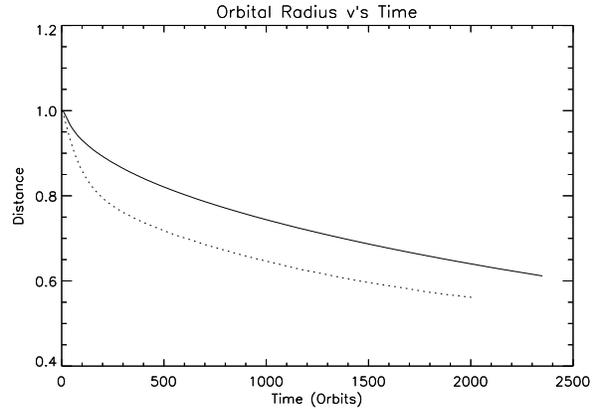,width=8.5cm}
\caption{\label{rpn4} This figure shows the evolution of the 
orbital radius for
the accreting protoplanet in run N3 (solid line) and for the non accreting 
protoplanet in run N4 (dotted line).}
\end{figure}


Calculation N3 was discussed in some detail in section~\ref{standard-run}.
Fig.~\ref{rpn4} shows the evolution of the orbital separations 
for calculations N3 (solid line) and N4 (dotted line). It is apparent
that the non accreting planet (N4) undergoes a significantly more rapid
phase of migration as the initial gap is cleared during the first few
hundred orbits, since the planet has to transfer a substantial amount of angular
momentum to the disc gas in order to clear the gap. 
The rapid accretion of mass (and
some angular momentum) by the accreting planet (N3)
helps to counteract the
initial inward torques that arise when the gap is being cleared of material,
as described in section~\ref{gap-no-gap}.

As the calculations proceed, the migration rates in both cases slow down
 when the gap has been cleared of material.  The planet  then migrates
inwards on approximately the viscous evolution time scale  of the outer disc. 
Some  additional slowing  down of the migration rates occurs because the protoplanets
interact with a smaller amount of disc mass as they migrate inwards. 

As the evolution time approaches $t=2000 \;P_0$, it is apparent that
the migration rate of the non accreting planet is actually slower
than that of the accreting planet, even though the accretion of
material has increased the inertia of the accreting planet.
The reason for this unexpected behaviour is that the non accreting planet
undergoes Roche lobe overflow. As material accumulates onto
the non accreting planet, the Roche lobe eventually
becomes filled with a hydrostatic atmosphere  and no additional
material may enter it. 
The continued flow of material from the outer
disc onto the protoplanet then leads to circulation
around it and  Roche lobe overflow, such that
material  flows towards the central star.

This material  contains too much angular momentum to flow through the
inner boundary directly, and instead fuels  the inner disc. This inner disc
then exerts a positive torque on the planet and reduces the rate at which it is
able to migrate towards the central star. This process provides an
efficient method
of allowing material to flow across tidally induced gaps in accretion discs,
and thus for the outer disc to feed material in to the inner disc which can
continue to accrete onto the central star. 
This process will be the subject of a more detailed future study. 

\begin{figure}
 \epsfig{file=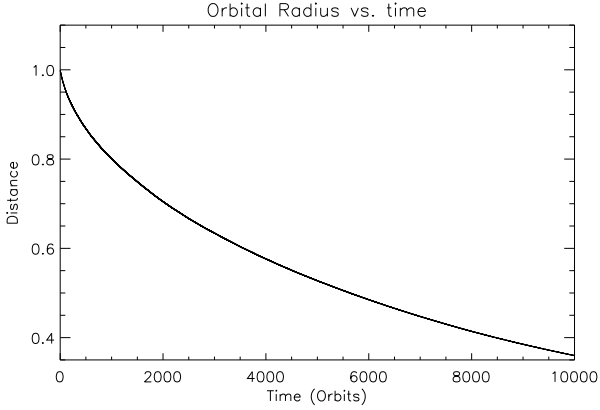,width=\columnwidth}
  \caption{This diagram shows the 
distance between the protoplanet and the primary as a function of time in
run F6.
At $t=0$ the gap is already cleared (it was  cleared during
400 fixed circular orbits).}
  \label{fig:ulmig}
\end{figure}

\subsection{ A Long-term Evolution Run}
The run F6 is a low/mid-resolution run aimed at computing the 
long-term behaviour of the accretion/migration process. As indicated in
table~\ref{tab1},  $N_r=70$
and $N_\phi=180$. Contrary to the other FARGO and NIRVANA runs, 
the inner boundary
is located at $r=0.25$ instead of $r=0.4$, and the radial grid spacing
is  in geometric  progression rather than uniform. 
Since we want to deal with the long-term behaviour of the protoplanet,
 we allow it to get close to the primary which is why we take a
smaller inner boundary radius. Furthermore, we want to track the
accretion rate as accurately as possible.  Taking  the non uniform
radial grid spacing ensures that the  cells all have  the same shape
and that the accretion algorithm will not be biased accordingly.  
For this run we first clear the gap with the inner boundary open
by evolving the  system  with the protoplanet orbit circular and  fixed
for 400 periods.
Then we start the accretion/migration process. Time $t=0$ is taken then.


In this run,  as distinct from the others, the frame is centered on the
centre of mass of the system composed of the primary and the protoplanet. 
 This is not an
inertial frame, since that would need to be centered on the centre of
mass of the primary, protoplanet and disc. But the indirect
term arising from the acceleration of this frame is much smaller than
the indirect term in the case of a frame centered on the primary.
Furthermore, the material in the outer disk tends to orbit around
the centre of mass of the primary and protoplanet (since the disk
itself is not self-gravitating), so one can work with a rigid
outer boundary and impose there a fixed Keplerian velocity,
with no radial motion, which avoids inflow and outflow
at that boundary.  
On the other hand, the material in the inner disk tends
to orbit around the primary, so there is a mismatch there between
the grid boundary and the gas orbits, which leads to a ``vacuum-cleaner''
effect  which drains the inner disk faster than a frame centered on the
primary would do. However, tests have shown that
the flux of mass at the inner boundary   is at
most 10 to  20~\% larger as a result of this effect. 

\begin{figure}
    \epsfig{file=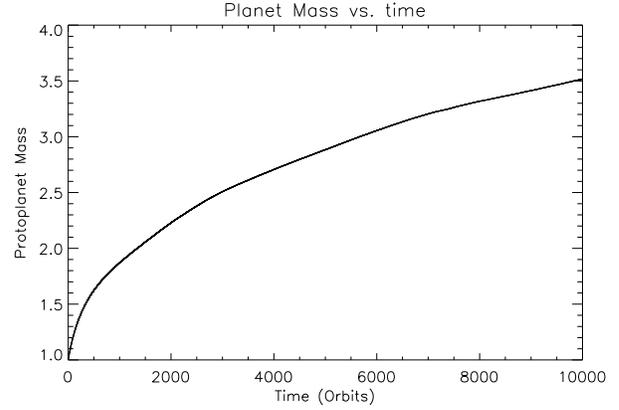,width=\columnwidth}
    \caption{This figure shows the temporal behaviour of the protoplanet
mass as a function of time in run F6. 
We notice that the curve slope (i.e. the
accretion rate) is not very peaked at $t=0$ since the gap was already
present. The mass unit is one Jupiter mass.}
    \label{fig:ulacc}
\end{figure}

We present in figure~\ref{fig:ulmig} the evolution of the 
protoplanet-primary separation as a function of time. In 
figure~\ref{fig:ulacc} we show the total mass of the protoplanet as 
a function of time, and in table~\ref{tab1} we give the estimated
migration time and final masses. These rates have been
extrapolated from the time derivatives of the  mass and orbital radius 
at time $t=7500$. This  corresponds roughly
to the time at which the inner disk is lost. 
The fact that these results are in relatively good agreement with the 
previous ones, even 
though the extrapolation is performed after many
more orbits in the case of run F6, shows that assuming that 
the migration and accretion
rates reach constant values is a reasonable approximation, even  though the
curves show some residual deviation from linearity.

\begin{figure*}
    \epsfig{file=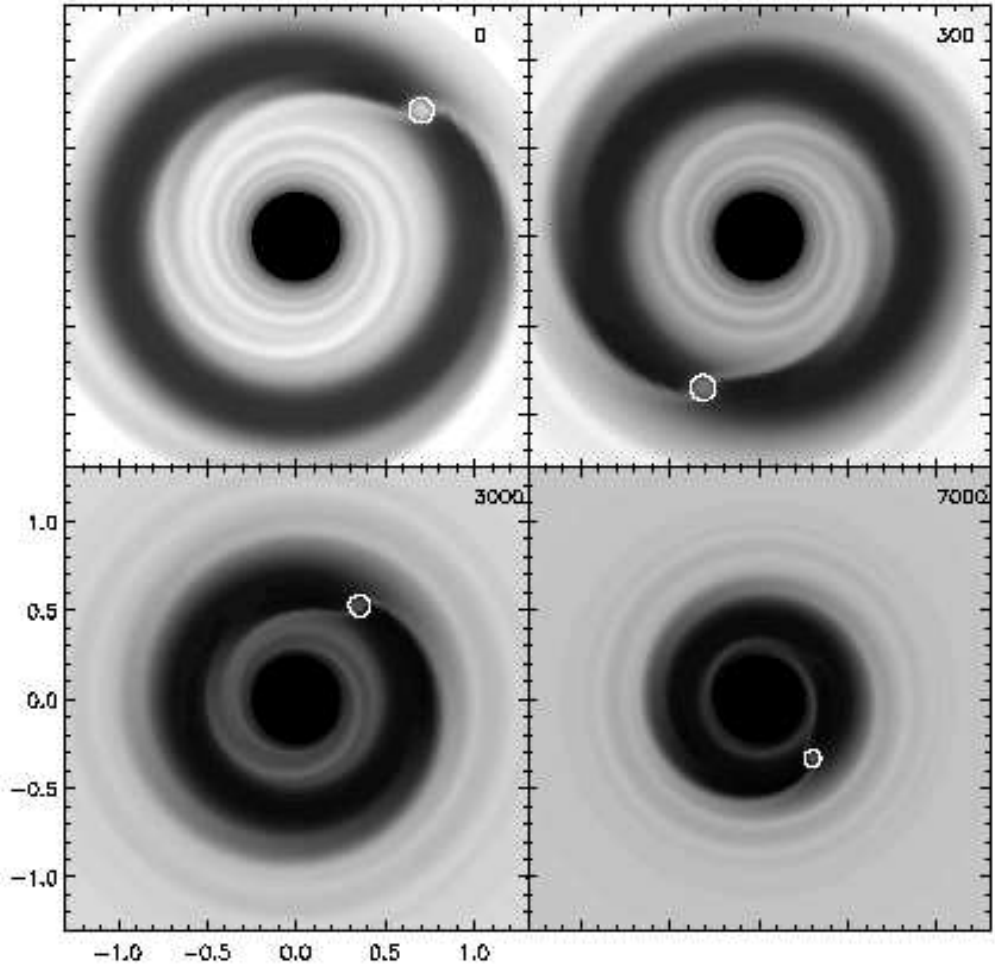,width=\textwidth}
    \caption{Time sequence of the surface density for run~F6. The outer
boundary is at $2.5$ and does not appear on the plots. The circle around
the protoplanet in each case has a radius equal to the Roche radius.}
    \label{fig:ulseq}
\end{figure*}

We show in figure~\ref{fig:ulseq} a sequence of four surface
density plots at  times ($t=0$, $t=300$, $t=3000$, $t=7000$).
As a
consequence of the migration  the gap radius decreases with
time. We  note as well the depletion of the inner disc (the
viscous time scale at $r=0.7$ is $\tau_{\rm visc}=7.8\cdot 10^3$~orbits,
and  is even smaller at smaller radii).

\begin{figure}
    \epsfig{file=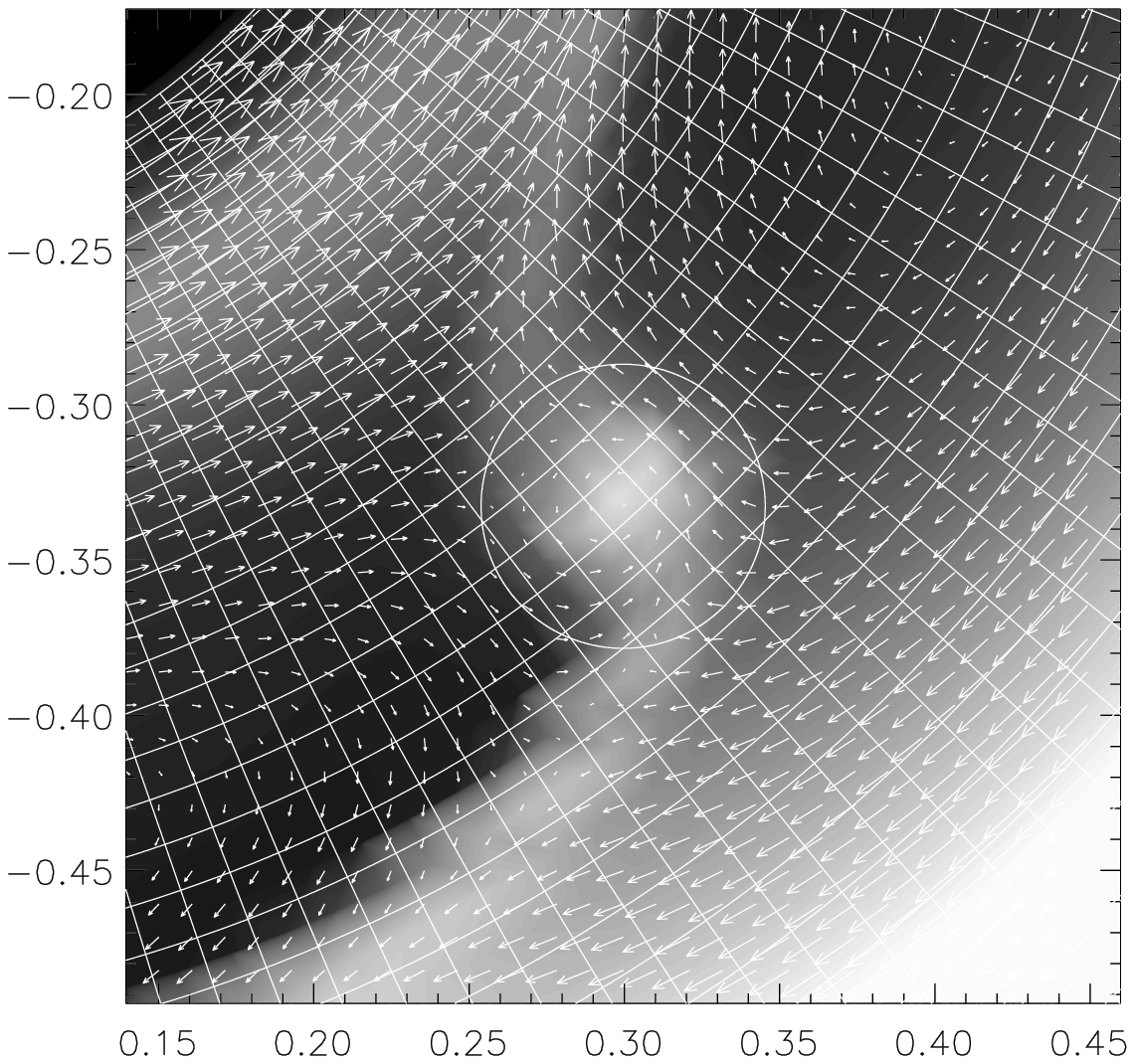,width=\columnwidth}
    \caption{Details of the flow in the neighbourhood of the protoplanet
near the end of the run F6 ($t=7000P_0$). The velocity field is represented
in the frame corotating with the planet.
We clearly see the gap, the wakes of the protoplanet in the inner and
outer disk, the extremities of the horseshoe orbits in the gap, 
the two X-points in the velocity field, which correspond to the
Lagrange points L$_1$ and L$_2$. The circle
is centered on the protoplanet and has a radius equal to the Roche
radius of the protoplanet. Because the actual potential felt by the
disk material is smoothed, the X-points lie slightly inside this
circle. One can see that the density peak around the planet is 
not elongated along the orbit, as usually observed in fixed frame
normal advection scheme runs. Because of the geometric grid spacing,
all the cells have the same shape and are ``as square as possible'',
i.e. $\log (r_{out}/r_{in}) \simeq N_r\log(1+2\pi/N_\phi)$.}
    \label{fig:detailend}
\end{figure}

\begin{figure}
    \epsfig{file=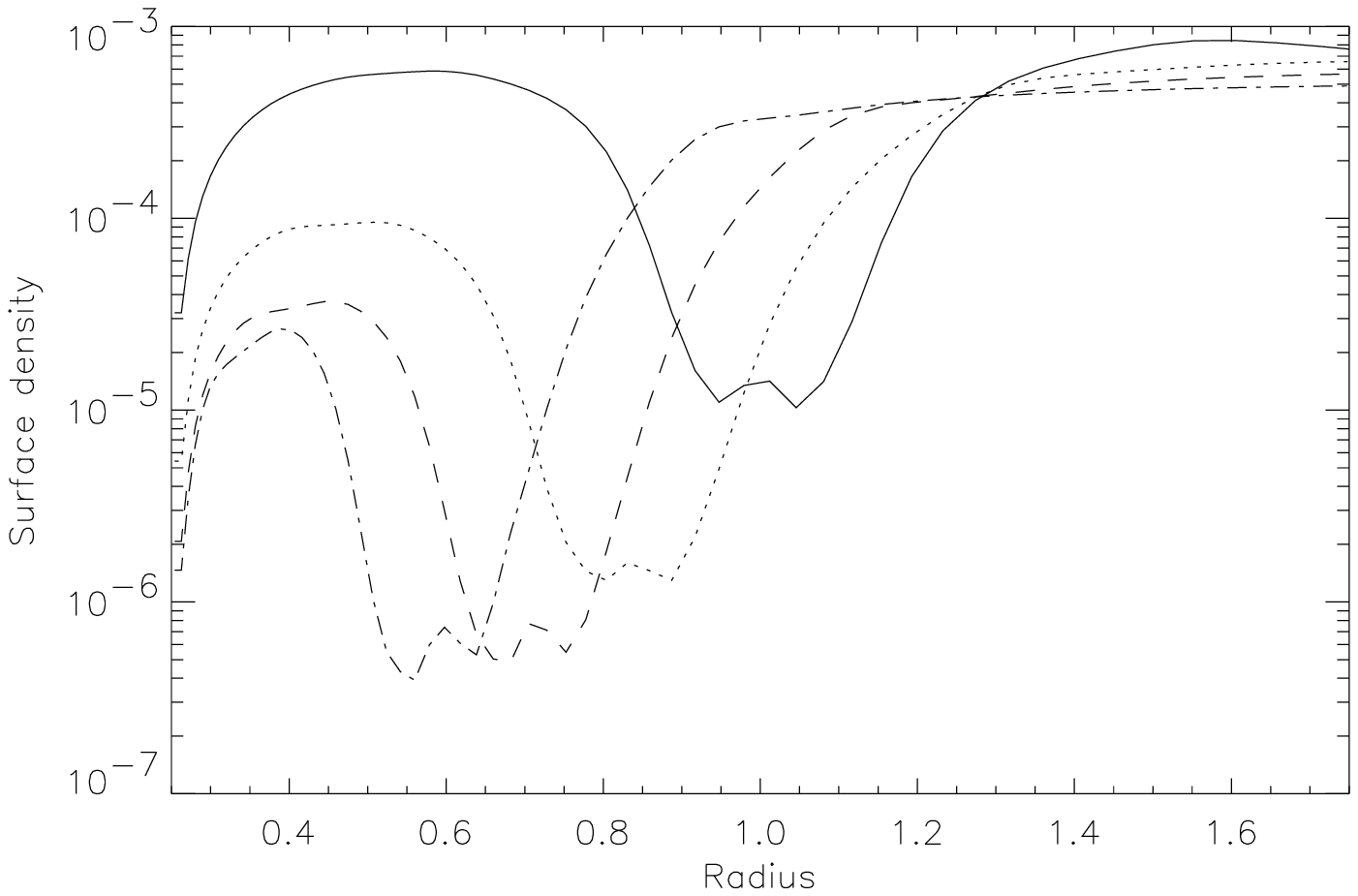,width=\columnwidth}
    \caption{Disk surface density profile at different times for run
F6 ($t=0$ : solid line, $t=690$ : dotted line, $t=1830$ : dashed line
and $t=3550$ : dot-dashed line). These plots represent the
azimuthal average of the surface density, and hence take the protoplanet
wake into account. The residual surface density which would be obtained
by omitting   the wakes  would be much smaller.}
    \label{fig:deepgap}
\end{figure}

Fig.~\ref{fig:detailend} shows the flow around the protoplanet
at the end of the run. Even though the inner disk is strongly
depleted, some Roche lobe overflow into it is indicated.
The mean profile of the gap surface density  at different times
is displayed in 
fig.~\ref{fig:deepgap}. The  gap deepens
as the protoplanet mass increases with time. The depletion of the inner disc
is also apparent.

\begin{figure}
    \epsfig{file=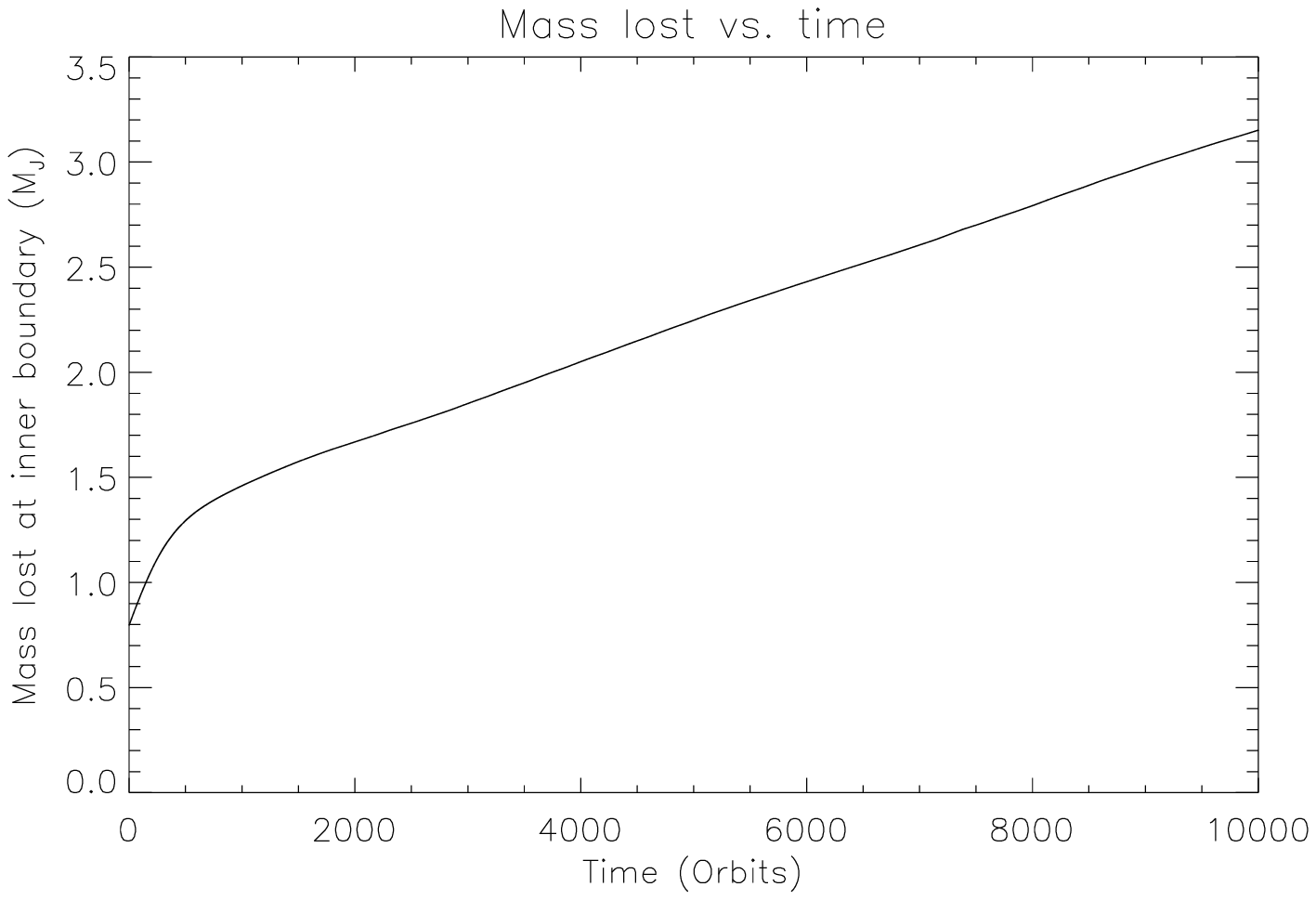,width=\columnwidth}
\caption{Mass lost at the inner boundary as a function of time in run F6. At
 $t=0$ this mass is not  zero since the run was   begun with
a non-accreting protoplanet on a fixed circular orbit with an open inner
boundary in order to  generate an initial gap.}
    \label{fig:lostmass}
\end{figure}

 The mass lost through the inner boundary
as a function of time is plotted in  fig~\ref{fig:lostmass}.
From the comparison of figs.~\ref{fig:ulacc}
and~\ref{fig:lostmass}, 
one can see that the mass overflow flux  
and
the mass accretion rate onto the planet have the same order of magnitude
(the mass overflow flux is of the same order of
magnitude as the mass outflow through
the inner boundary since the mass lost at the inner boundary is bigger
than the inner disk mass; in a totally stationary case these two rates
would be strictly equal).

The value of the ratio of the mass flow rate  through
the gap to the accretion rate
 $\sim 1/3.5$ 
obtained here
is  somewhat higher than  that obtained from the simulation $R2$ where
this ratio 
was about $1/7.$  However, the  magnitude of the accretion rate
 is smaller here since the planet mass
is larger.  
  Also the surface density  
profiles  and disk aspect ratios are different in these simulations. 
Given the differing physical situation when the accretion
rates were measured, we consider the agreement to be satisfactory.

\begin{figure}
\epsfig{file=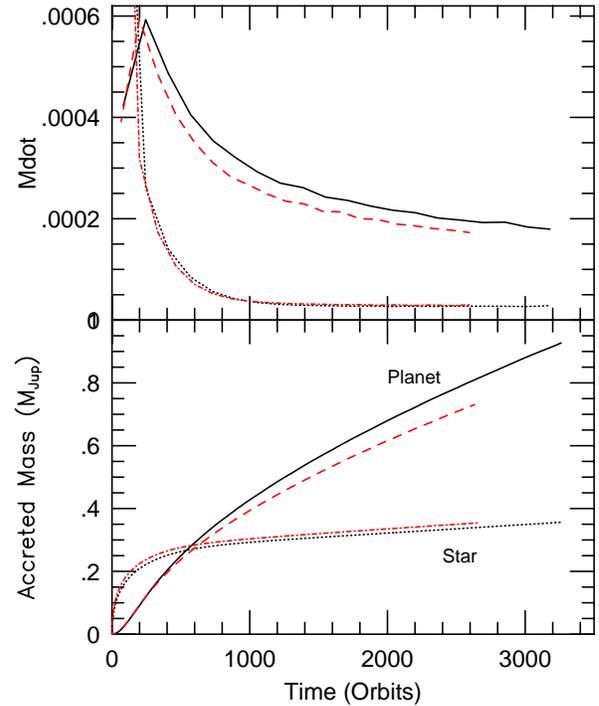, width=8.5cm}
\caption{\label{wk1} 
Mass evolution for models $R2$ and $R3$:
The lower panel shows the evolution of the mass of the planet and
the mass lost from the disc (accreted by the star).
The top panel shows the corresponding
accretion rates in units of $\MJup/P_0$.
Dark solid and dotted lines refer to model R2, while the lighter dashed and
dashed-dotted lines refer to the higher resolution model R3.}
\end{figure}

\subsection{ Runs with an Initial Surface Density Profile Using RH2D } 
\label{result:rh2d}
The code RH2D has been already been tested against NIRVANA (Kley 1998)
and found to give very similar results so we shall
not give results of additional tests here.
Instead we present runs R$i$ ($i=1,2,3$) which 
use different  initial conditions
incorporating a surface density profile that is not constant,
as outlined at the end
of section \ref{ini-bc}, and a slightly higher value of the disc aspect ratio
$H/r=0.05$. Also, it should be noted that these calculations are initiated
with a gap already in existence around the position of the protoplanet.

The main results concerning the mass accretion by the protoplanet are
displayed in Fig.~\ref{wk1}. The lower panel
shows the evolution of the protoplanet mass and the mass lost through
the inner boundary for the  models R2 and R3 which have
maximal accretion. These models have the same
physical setup but different numerical resolutions.
Note that only the mass added to the planet during the evolution
is displayed. The total planet mass is obtained by adding  $1M_{J}$ to
the quoted values, {\em i.e.} at the end of the run R2 at $t \approx 3300$
the planet has reached about  $2M_{J}$. This value is smaller
than that obtained by the other codes (see for example run F6),
which is a result of 
the surface density near the planet being a factor of about three smaller
in  these runs.
Also shown in the lower panel is the mass lost through the inner boundary
(dotted and dashed dotted lines)
which is assumed to have been accreted by the star.
The mass of the planet rises more slowly initially, 
because of the initial gap imposed.
Only after the full development of the quasi-stationary flow
(at $t \approx 300$) does the
mass accumulation rate onto the planet become larger than the mass loss rate
through the inner boundary.
This is demonstrated in the top panel where the mass accretion rates
in units of $M_J/P_0$
onto the planet and star are shown.
After the mass contained initially within $r=0.25$ and $0.4$,
$M_{d0}=0.37 \MJup$
is consumed (primarily by the star), the mass accretion rate onto the star
settles to the very small constant value $\dot{M}_{lost}= 2.86 \times 10^{-5}
\MJup/P_0$
which may be identified with the
rate of mass overflow across the gap.
Near the end of the computation at $t=3000$ this rate is about 6.5 times lower
than the mass accretion rate by the planet which is in rather good
agreement with the results quoted in Kley (1999a) but somewhat lower
than the rates obtained with FARGO as outlined above.

\begin{figure}
\epsfig{file=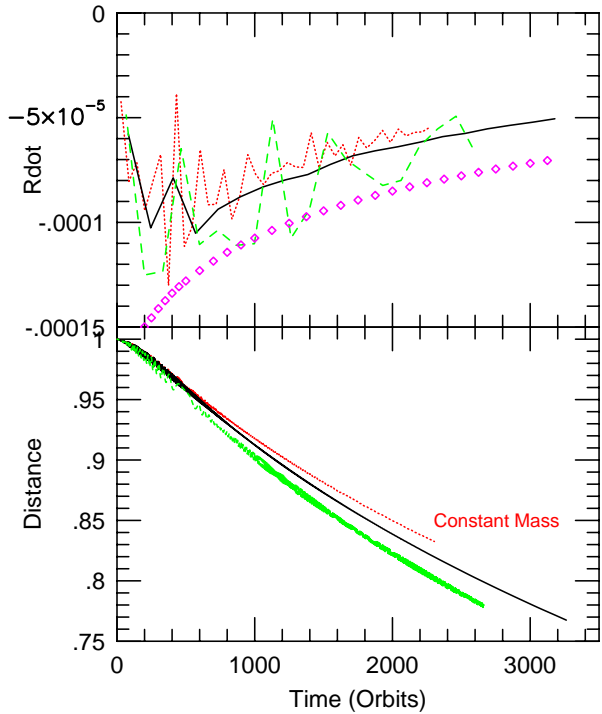, width=8.5cm}
\caption{\label{wk2}
Radial evolution for models R$i$:
The lower panel shows the evolution of the semi major axes of the
planet for three different models. Solid black line R2, dotted grey
line the non-accreting model R1, and dashed light grey line the
high resolution model R3. The top panel shows the corresponding
derivative $\dot{r}_p$ in units of $a_0/P_0$ versus time.
The diamonds represent the analytical
approximation according to Ivanov, Papaloizou, \& Polnarev  (1999)
as given in Eq.~\ref{rdot-Ivanov} for model R2.}
\end{figure}

In Fig.~\ref{wk2} the evolution of the distance of the planet from the 
star is shown for all models R1--R3. Again, due to the smaller mass
contained in the physical domain the orbital decay rate is smaller than
in the previous models (N$i$ and F$i$). In the comparison
model labeled `constant
mass' (R1, dotted line)
the mass of the planet remained constant but mass was nevertheless
removed from the Roche--lobe in the manner described in section \ref{accretion}
(also see Kley, 1999a).  
This model has a smaller decay rate than the complete models (R2 solid line,
and R3, light dashed line) in
which the accreted mass is added to the planet mass. 
All the models show initially a fluctuation in $r_p(t)$ which indicates
an eccentricity growth to less than about 0.02
which damps out at later times.
For the high resolution model (R3, light dashed line) the radial evolution
follows closely the low resolution model, the migration rate seems to
be marginally lower. 
This is consistent with the results obtained with NIRVANA and FARGO
which indicate that lower resolution runs have larger accretion rates
and slower decay rates (see Figs.~\ref{rpn7} and \ref{fig8}).

The diamonds in Fig.~\ref{wk2} represent the analytical
approximation according to Ivanov, Papaloizou, \& Polnarev (1999)
as given in Eq.~\ref{rdot-Ivanov}
with an arbitrary  normalisation factor which can be adjusted 
to match the numerically obtained data.
For simplicity here the maximum of the azimuthally averaged surface density
$\Sigma(r)$ outside of the planet was
taken for model R2 as the value for the surface density in (\ref{rdot-Ivanov}). 
Clearly the formula gives an excellent approximation to the actual evolution
of the radius of the planet. 

\subsection{Comparison between NIRVANA and FARGO} \label{compare-codes}

\begin{figure*} 
\epsfig{file=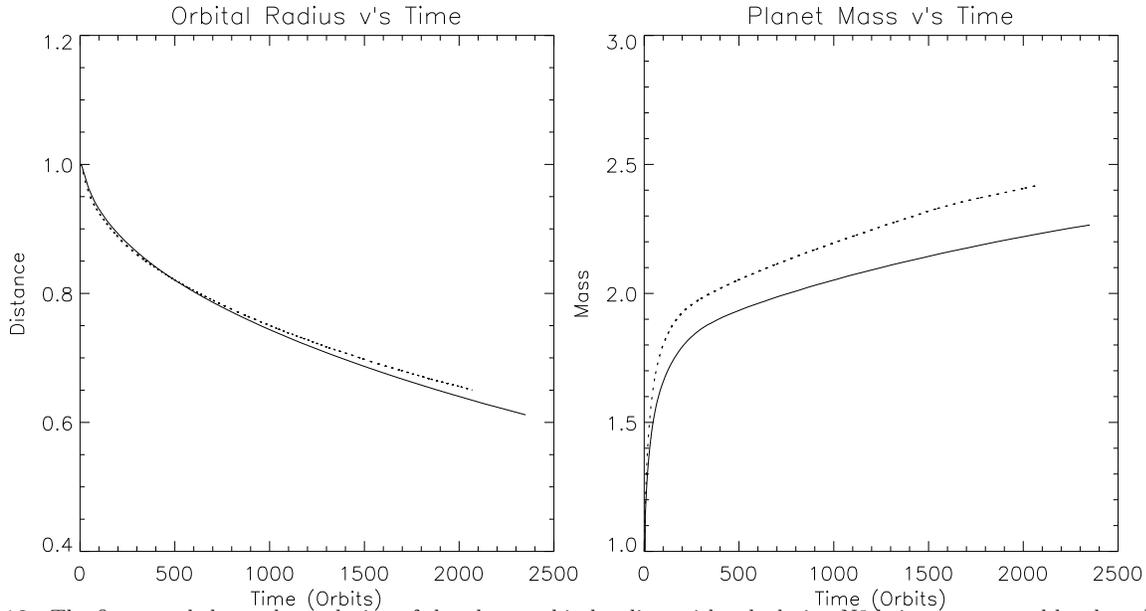, width=15cm}
\caption{\label{rpn5} The first panel shows the evolution of the planet orbital radius,
with calculation N3 being represented by the solid line and F3
by the dotted line.
The second panel shows the increase in mass as the planet accretes
gas from the disc, with the same line styles representing the two calculations.}
\end{figure*}

In this section we present the results of calculations that were
performed with both NIRVANA and FARGO in order to check that
the results that we have presented are reproducible when using 
independent numerical codes. The calculations that we will
compare are N3 performed using NIRVANA and F3 performed using FARGO.
As described in table~\ref{tab1}, these calculations are for an accreting 
Jupiter mass planet initially embedded in an unperturbed disc model.
For comparison purposes, the orbital evolution of the protoplanet is 
plotted in the first panel of fig.~\ref{rpn5}, with calculation N3 being
represented by the solid line and F3 being represented by the dotted line.
Similarly, the evolution of the protoplanet mass is presented in the second
panel of this figure with the same line styles being used to represent the
calculations. It is apparent that the results are very closely matched in terms
of the orbital evolution rate and reasonably well matched in terms
of the mass accretion rates.
We clearly see the trend, also seen  in the results
given in table~\ref{tab1},  that  the FARGO algorithm
 gives  higher accretion rates 
 and hence 
slightly slower migration.
This may be related to the fact that  the distribution of matter
close to the protoplanet is subject to less numerical diffusion
and so has less azimuthal elongation in FARGO.

\subsection{Numerical Resolution} \label{resolution}
\subsubsection{Non accreting planets}

In this section we present the results of calculations that were
performed using different resolutions. 
We  first concentrate on
comparing three calculations
in which the orbital evolution of a non accreting planet was studied, 
namely calculations N2, N4, and N6.

\begin{figure} 
\epsfig{file=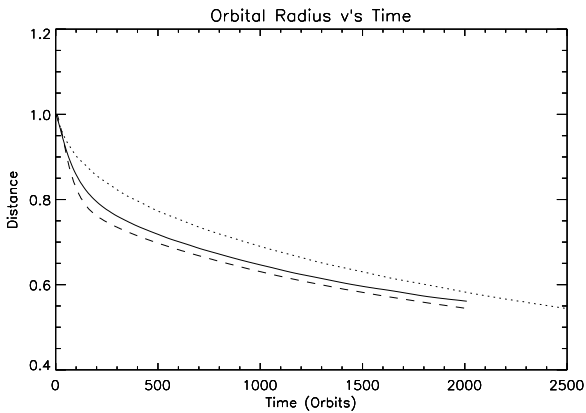, width=8.5cm}
\caption{\label{rpn6} This figure shows the evolution 
of the orbital radius for the
calculations N2 (dotted line), N4 (solid line), and N6 (dashed line).
The resolution for each calculation is described in table~\ref{tab1}.}
\end{figure}

The evolution of the planet orbit radius for these calculations
is presented in fig.~\ref{rpn6}, with calculation N2 being represented
by the dotted line, calculation N4 by the solid line, and calculation N6 
by the dashed line. The agreement between N2 and the other calculations appears
to be the worst, which is to be expected since this
is the lowest resolution simulation that we performed and this  is probably too
low to fully resolve small scale structures in the vicinity of the planet.
The two calculations N4 and N6 on the other hand show extremely good agreement
in their orbital migration rates at later times, though there is a small offset
in the orbital radius at any given time due to calculation
N6 experiencing more rapid
migration during the gap clearing stage of the calculation. 

This 
increased rate of orbital migration for the higher resolution calculation
during the gap clearing stage probably arises because of its ability
to resolve density waves with higher azimuthal mode numbers that are located
close to the planet when it is embedded. Once the gap has been cleared,
however, these non axisymmetric structures do not provide a significant
contribution to the tidal torque acting on the planet, so that the 
migration rates then become approximately equal. The close agreement
between the migration rates of calculations N4 and N6 indicate that
our calculations have essentially reached convergence in their results.

\subsubsection{Accreting planets}

\begin{figure*}
\epsfig{file=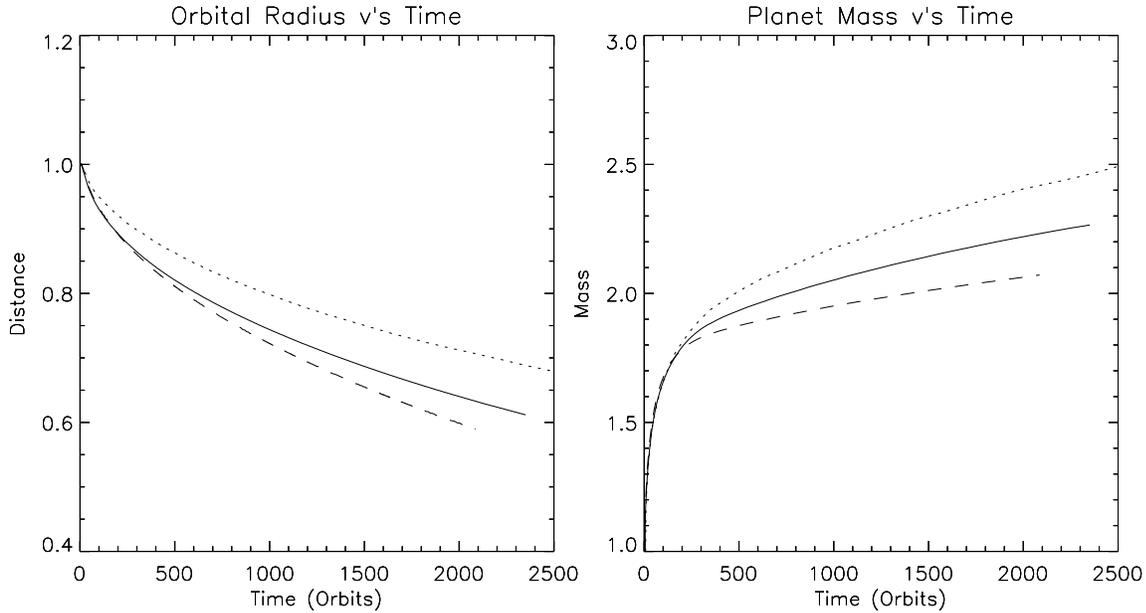, width=15cm}

\medskip

\caption{\label{rpn7} The first panel shows the evolution of the orbital radius for the
calculations N1 (dotted line), N3 (solid line), and N5 (dashed line). The
evolution of the planet mass is shown in the second panel, with the
line styles being the same as above.
The resolution for each calculation is described in table~\ref{tab1}.}
\end{figure*}

\begin{figure*}
\epsfig{file=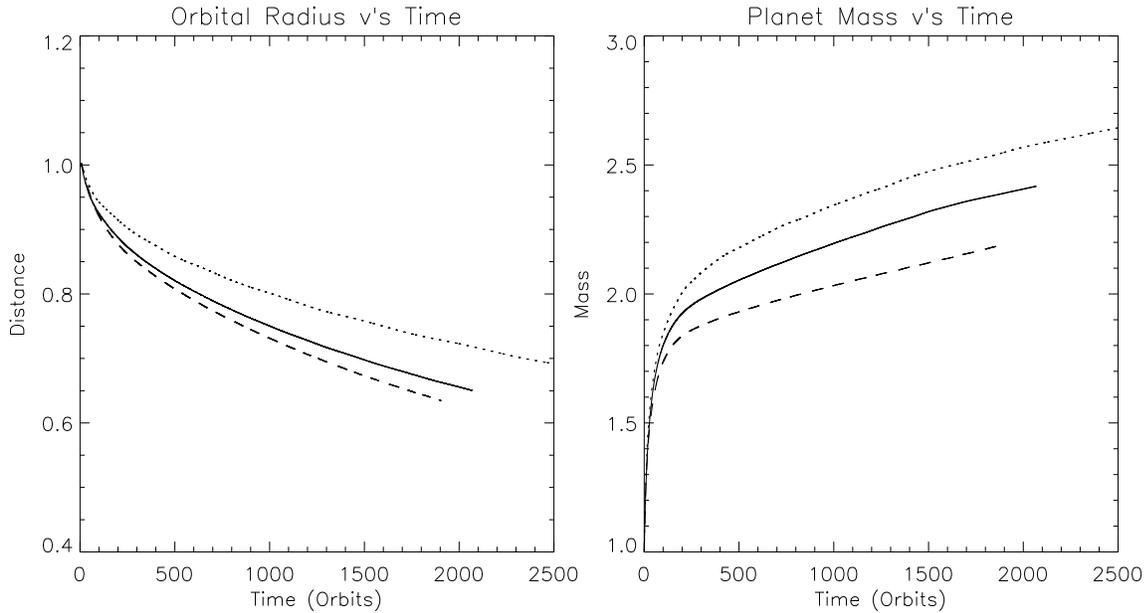, width=15cm}

\medskip

\caption{\label{fig8} Same as figure~\ref{rpn7} for FARGO results F1 (dotted line),
F3 (solid line) and F5 (dashed line).}
\end{figure*}

The calculations for accreting planets initially embedded in unperturbed
accretion discs that were performed with different resolutions
are shown in figs.~\ref{rpn7} and~\ref{fig8}. The first panel in 
fig.~\ref{rpn7} shows the
orbital radius as a function of time for the calculations
N1 (dotted line), N3 (solid line), and N5 (dashed line).
The second panel shows the evolution of the planet mass, with the
same line styles as described for the first panel. 

Fig.~\ref{fig8}
 gives the corresponding results for the runs F1, F3 and F5.
It is apparent that the lower resolution
runs have a  larger accretion rate.  
This  leads to  smaller migration rates because
the protoplanets have larger masses.
 Nonetheless, the migration times
obtained for these runs all indicate an orbital decay time of $\tau_{mig} \sim
10^4 \;P_0$, as indicated in table~\ref{tab1}. This is in agreement
with the idea that the orbits of
giant planets will evolve on the viscous diffusion time of 
the protostellar disc.
The results of NIRVANA and FARGO are in good agreement
with respect to migration rates and final planetary masses.
The predicted final
planetary masses in the calculations N3 and N5 are  $3.2 \MJup$ and 
$2.7 \MJup$  while for the
calculations F3 and F5 they are $3.85 \MJup$ and $3.25 \MJup$ respectively.
These  agree well with the estimated masses of a
number of the recently discovered
closely orbiting extrasolar giant planets (Marcy, Cochran, \&  Mayor 1999,
Marcy \& Butler 1998).

\section{Discussion and Conclusion} \label{conclusion}

In this paper we have studied the interaction of a protoplanet
of $1\MJup$ initially with a gaseous disc whose aspect ratio and kinematic
viscosity are those expected for a minimum mass solar nebula.
This had characteristically $2\MJup$ interior to the initial circular orbit
radius of the protoplanet.
The problem was studied with three independent
hydrodynamic codes, NIRVANA, FARGO
and RH2D. These were found to give consistent results when compared.
FARGO had the additional advantage that, on account of the fast
advection scheme employed, the evolution could be followed
for a much longer time.

A general result of the simulations was that the direction
of the orbital migration was always inwards and such that the
protoplanet reached the central star in a near circular orbit
after a time of $\sim 10^4$
initial orbital periods, which is characteristically
the viscous time scale at the initial orbital radius.
This was found to be independent
of whether the protoplanet was allowed to accrete mass or not,
or the surface density profile in the disc. The tendency to migrate
inwards was assisted by the disappearance of the inner disc
through accretion onto the central star.
When the protoplanet was allowed to accrete at a near maximal rate
the mass was found to increase to about $4\MJup$ 
as it reached the central star. Because of deep gap
formation and lower accretion rates for the larger masses (also see BCLNP)
it is difficult to exceed this mass in the kind of simulations
presented here. An additional calculation has been performed with an initial
planet mass of $3\MJup$, but whose results are not described here in detail.
This calculation resulted in a similar migration time
of $\sim 10^4 \; P_0$, and an estimated final mass for the protoplanet of
$4.8 \; \MJup$, indicating the difficulty of forming giant planets
with masses greater than about 5 $\MJup$ before they have migrated to the
centre.
It would appear that the masses estimated for a number of close orbiting
giant planets  (Marcy, Cochran, \& Mayor 1999, Marcy \& Butler 1998) 
as well as their inward orbital migration can be explained
by the  operation of the processes considered here
during the late
stages of giant planet formation.  

Several important issues, however, remain to be resolved.
The inward migration time is shorter than previously estimated
time scales $\sim 10^6$~yr for the formation of a Jovian mass protoplanet
(e.g. Bodenheimer \& Pollack 1986; Pollack {\em et al}. 1996; 
Papaloizou \& Terquem 1999).
This is suggestive that either the planet formation
process may be speeded up by the earlier
merger of cores undergoing type I migration, and/or there may be regions
in the disc where the viscosity is very small, so halting
type II migration,  perhaps due to inadequate
ionization for MHD instabilities to operate. 
Additional planets embedded in the disc 
alter the density structure and consequently the torque balance
which may result in a halting of the migration process (Kley 1999b). 
Many body processes
such as gravitational
scattering of protoplanets may also operate to move them to different
orbital locations in the disc (e.g. Weidenschilling \& Marzari 1996). 

An additional possibility is that in some cases
giant planet formation occurs at substantially larger
distances from the host star than have hitherto been given serious
consideration. For example, a planet forming at a radius $\sim 20$ AU
will have a migration time of $\gs \; 10^6$ yr, which is now within the
estimated range of lifetimes of protostellar discs around T Tauri 
stars ({\em i.e.}
$10^6$ -- $10^7$ yr). It is possible that inward migration of these
protoplanets may simply be halted by the eventual dissipation of the
disc at the end of the T Tauri stage. The final orbital
positions of these planets will then be determined by the initial 
radius at which the
planets were formed and the age at which the T Tauri phase ends. In this
scenario, planets that initially start to form closer in towards the
central star (e.g. at 5 AU) will migrate inwards and will become `hot Jupiters',
whereas those planets that form further away 
stand a much greater chance of being at intermediate distances from their
host stars when orbital migration is halted by the disappearance of the disc.

Another issue is that type II migration in a viscous disc as considered
here tends to cause Jovian mass protoplanets to merge with their
central star on a time scale short compared to the lifetime of
protostellar discs. Thus a process for halting the migration
is required.  This may occur through the termination
of the inner disc due to a magnetospheric cavity 
(Lin, Bodenheimer, \& Richardson 1996).

The calculations presented here make a number of important assumptions that
may have some bearing on the final results obtained.
By using a locally isothermal equation of state, we tacitly assume that
any heat generated by the spiral shock waves is immediately radiated from the 
system. This may not be an accurate description of the
thermodynamics, and if radiative processes operate on a time scale
that is longer than the orbital time scale then some thickening of the
disc may result. In addition, we assume that the turbulent viscosity can be
simply modeled using the Navier-Stokes equation, when in reality it should
arise naturally from MHD instabilities. These, and other assumptions
can only be addressed by performing global simulations that
include radiative and MHD processes in three dimensions. We hope in the near
future to be able to address some of these outstanding issues.

\section{Acknowledgements}
This work was supported in part (F.M.) by the European
Commission under contract number ERBFMRX-CT98-0195
(TMR network ``Accretion onto black holes, compact stars 
and protostars''), and by the Max-Planck-Gesellschaft,
Grant No. 02160-361-TG74 (for W.K.).
Computational resources of the Max-Planck-Institute for Astronomy
in Heidelberg were available and are gratefully acknowledged. 
We thank Udo Ziegler for making a FORTRAN Version of his code NIRVANA
publicly available.

\end{document}